\documentclass[12pt,preprint]{aastex}

\newcommand{\chandra}{{\emph{Chandra}}}

\newcommand{\sst}{{\emph{Spitzer Space Telescope}}}

\newcommand{\beqarr}{\begin{eqnarray}}
\newcommand{\eeqarr}{\end{eqnarray}}
\newcommand{\beq}{\begin{equation}}
\newcommand{\eeq}{\end{equation}}

\def\simlt{\mathrel{\spose{\lower 3pt\hbox{$\mathchar"218$}}
     \raise 2.0pt\hbox{$\mathchar"13C$}}}
\def\simgt{\mathrel{\spose{\lower 3pt\hbox{$\mathchar"218$}}
     \raise 2.0pt\hbox{$\mathchar"13E$}}}

\shorttitle{Mid-IR Variability from the {\it Spitzer} Deep Wide-Field Survey}
\shortauthors{Koz{\l}owski et al.}

\begin{document}


\title{Mid-Infrared Variability from the {\it Spitzer} Deep Wide-Field Survey}

\author{
Szymon~Koz{\l}owski\altaffilmark{1},
Christopher~S.~Kochanek\altaffilmark{1,2},
Daniel~Stern\altaffilmark{3},
Matthew~L.~N.~Ashby\altaffilmark{4},
Roberto~J.~Assef\altaffilmark{1},
J.~J.~Bock\altaffilmark{5},
C.~Borys\altaffilmark{5},
K.~Brand\altaffilmark{6},
M.~Brodwin\altaffilmark{4,21},
M.~J.~I.~Brown\altaffilmark{7},
R.~Cool\altaffilmark{8},
A.~Cooray\altaffilmark{9},
S.~Croft\altaffilmark{10},
Arjun~Dey\altaffilmark{11},
P.~R.~Eisenhardt\altaffilmark{3},
A.~Gonzalez\altaffilmark{12},
V.~Gorjian\altaffilmark{3},
R.~Griffith\altaffilmark{3},
N.~Grogin\altaffilmark{6},
R.~Ivison\altaffilmark{13,14},
J.~Jacob\altaffilmark{3},
B.~T.~Jannuzi\altaffilmark{11},
A.~Mainzer\altaffilmark{3},
L.~Moustakas\altaffilmark{3},
H.~R\"ottgering\altaffilmark{15},
N.~Seymour\altaffilmark{16},
H.~A.~Smith\altaffilmark{4},
S.~A.~Stanford\altaffilmark{17},
J.~R.~Stauffer\altaffilmark{18},
I.~S.~Sullivan\altaffilmark{6},
W.~van Breugel\altaffilmark{19},
S.~P.~Willner\altaffilmark{4}, 
and 
E.~L.~Wright\altaffilmark{20}\\
}

\altaffiltext{1}{Department of Astronomy, The Ohio State University, 140 West 18th Avenue, Columbus, OH 43210, USA; e-mail:  
{\tt simkoz@astronomy.ohio-state.edu}}
\altaffiltext{2}{The Center for Cosmology and Astroparticle Physics, The Ohio State University, Columbus, OH 43210, USA}
\altaffiltext{3}{Jet Propulsion Laboratory, California Institute of Technology, 4800 Oak Drive, Pasadena, CA 91109, USA}
\altaffiltext{4}{Harvard-Smithsonian Center for Astrophysics, 60 Garden Street, Cambridge, MA 02138, USA}
\altaffiltext{5}{California Institute of Technology, Pasadena, CA 91125, USA}
\altaffiltext{6}{Space Telescope Science Institute, Baltimore, MD 21218, USA}
\altaffiltext{7}{School of Physics, Monash University, Clayton 3800, Victoria, Australia}
\altaffiltext{8}{Department of Astrophysical Sciences, Princeton University, Princeton, NJ 08544, USA}
\altaffiltext{9}{University of California, Irvine, CA 92697,USA}
\altaffiltext{10}{University of California, Berkeley, CA 94720,USA}
\altaffiltext{11}{National Optical Astronomical Observatory, 950 North Cherry Avenue, Tucson, AZ 85719, USA}
\altaffiltext{12}{Department of Astronomy, University of Florida, Gainesville, FL 32611, USA}
\altaffiltext{13}{UK Astronomy Technology Centre, Royal Observatory, Blackford Hill, Edinburgh, EH9 3HJ, UK}
\altaffiltext{14}{Institute of Astronomy, University of Edinburgh, Blackford Hill, Edinburgh, EH9 3HJ, UK}
\altaffiltext{15}{Leiden Observatory, Leiden University, PO Box 9513, 2300 RA Leiden, The Netherlands}
\altaffiltext{16}{Mullard Space Science Laboratory, University College London, UK}
\altaffiltext{17}{University of California, Davis, CA 95616, USA}
\altaffiltext{18}{Spitzer Science Center, California Institute of Technology, Pasadena, CA 91125, USA}
\altaffiltext{19}{University of California, Merced, CA 95344, USA}
\altaffiltext{20}{University of California, Los Angeles, CA 90095-1562, USA}
\altaffiltext{21}{W. M. Keck Postdoctoral Fellow at the Harvard-Smithsonian Center for Astrophysics}


\begin{abstract}
We use the multi-epoch, mid-infrared {\it Spitzer} Deep, Wide-Field Survey to investigate the variability 
of objects in 8.1 deg$^2$ of the NOAO Deep Wide Field Survey Bo{\"o}tes field. We perform a Difference Image 
Analysis of the four available epochs between 2004 and 2008, focusing on the deeper 3.6 and 4.5 \micron~bands.
Out of $474179$ analyzed sources, 1.1\% meet our standard variability selection criteria, that the two light curves are strongly correlated
($r>0.8$) and that their joint variance ($\sigma_{12}$) exceeds that for all sources with the same magnitude by $2 \sigma$. 
We then examine the mid-IR colors of the variable sources and match them with X-ray sources from the
XBo{\"o}tes survey, radio catalogs, 24 \micron\, selected active galactic nucleus (AGN) candidates, and spectroscopically
identified AGNs from the AGN and Galaxy Evolution Survey (AGES). Based on their mid-IR colors, 
most of the variable sources are AGNs (76\%), with smaller contributions from stars (11\%), galaxies (6\%), 
and unclassified objects, although most of the stellar, galaxy, and unclassified sources are false positives.
For our standard selection criteria, 11\%--12\% of the mid-IR counterparts to X-ray sources, 24 \micron~AGN candidates, and 
spectroscopically identified AGNs show variability. The exact fractions depend on both the search depth
and the selection criteria. For example, 12\% of the 1131 known $z>1$ AGNs in the field and 14\%--17\% of the known AGNs with well-measured fluxes
in all four IRAC bands meet our standard selection criteria.
The mid-IR AGN variability can be well described by a single 
power-law structure function with an index of $\gamma\approx0.5$ at both 3.6 and 4.5 \micron, and an amplitude of 
$S_0\simeq0.1$ mag on rest-frame timescales of 2 years. The variability amplitude is higher for shorter rest-frame wavelengths and
lower luminosities. 

\end{abstract}
\vspace{2.2cm}



\section{Introduction}
\label{sec:intro}

A substantial fraction of objects in the universe change in brightness with time. This apparent variability 
can be due to physical periodic changes in the objects (e.g., Cepheids, RR Lyrae), motion (e.g., eclipsing binaries, 
microlensing, rotating stars), explosive events (e.g., supernovae, novae), and accretion (e.g., cataclysmic variables, 
active galactic nuclei (AGNs)). In a high-latitude extragalactic survey, the most common variable sources are 
active galactic nuclei (e.g., \citealt{2007AJ....134.2236S}).  Optical variability is one method for 
identifying quasars\footnote{We will use words ``quasar'' and ``AGN'' interchangeably throughout this paper.} 
(e.g., \citealt{2002AcA....52..241E,2003AJ....125....1G,2004ApJ...606..741R}),
although it is only recently that a fully quantitative approach to variability selection has been developed
(\citealt{2010ApJ...708..927K}).

The emission from quasars generally has three components.  The UV to near-IR radiation is dominated by a
hot accretion disk extending from an inner edge of a few gravitational radii from the black hole outward
with, in simple thin disk theory, a temperature profile $T \propto R^{-3/4}$ (\citealt{1973A&A....24..337S}).  Near the inner edge,
there is a corona of hotter gas that produces the non-thermal X-ray emission (\citealt{1994ApJ...432L..95H}).  On scales where the 
temperature is below the dust sublimation temperature ($\sim 2000$~K), dust absorbs radiation from the
disk and reradiates the energy in the mid-IR and far-IR (\citealt{1987ApJ...320..537B}).  The overall spectrum typically shows a minimum
near 1 \micron, with the emission from the disk rising toward the UV and the emission reradiated by 
dust rising toward the far-IR (e.g., \citealt{1989ApJ...347...29S}). 
There is an increasing evidence that many physical relations between AGNs,
galaxies and their large scale clustering have to be taken into account
if their formation and evolution is to be understood (e.g., \citealt{2008A&A...490..893T}).

We know a great deal about the optical variability of quasars both from large studies of the variability
seen in ensembles of sparsely monitored quasars and from detailed studies of individual quasars.  Ensemble
studies (e.g., \citealt{2004ApJ...601..692V,2005AJ....129..615D}) 
have shown that variability increases with decreasing optical wavelength, decreasing luminosity,
and potentially decreasing black-hole mass.  The structure function of the ensemble variability is
a power law with smaller variability amplitudes on short timescales, with some evidence for saturation on timescales of order
a few decades.  Until recently, there were few studies of individual quasars (e.g.,
\citealt{1985ApJ...296..423C,1989ApJ...337..236C,1994MNRAS.268..305H}), but this has
changed dramatically in the last year.  The light curves of individual quasars are well modeled
by a damped random walk, a stochastic process described by the amplitude of the random walk and a
damping timescale for returning to the mean luminosity (\citealt{2009ApJ...698..895K,2010ApJ...708..927K,2010arXiv1004.0276M}).  
While preliminary indications from \cite{2009ApJ...698..895K}, based on $\sim100$ quasars, suggest that 
these two process parameters are related to the quasar luminosity 
and black hole mass, \cite{2010arXiv1004.0276M} used $\sim9000$ Sloan Digital Sky Survey (SDSS) Stripe 82 quasars to find a number of clear trends.
For example, the asymptotic variability on long timescales decreases with increasing luminosity and rest-frame wavelength, and is correlated
with black hole mass. The timescale for returning to the mean luminosity increases with wavelength and also with increasing black hole mass, but
remains constant with redshift and luminosity. 

Far less is known about the near-IR and mid-IR variability of quasars. Since AGN luminosities do vary in time at optical wavelengths, 
we expect to find some variability in the IR, but it could be heavily 
smoothed by averaging the response over the large scales of either the cooler parts of the disk or the absorbing dust.
\cite{2004MNRAS.350.1049G} found that the majority (39/41) of the surveyed low-luminosity nearby Seyferts do vary in the near-IR and mid-IR, and
 that the variability is most apparent at longer wavelengths due to diminishing flux from the host galaxy.
The most extensive work to date on near-IR and mid-IR
variability of quasars is that of \cite{1999AJ....118...35N}. These authors studied 25 low-redshift quasars ($z\simeq0.1$), 
which were observed in five bands (1.3 -- 10.6 \micron) for $\sim30$ years. The amplitude of variations 
for radio loud (quiet) quasars is $\sim0.3$ mag (0.1 mag)
on rest-frame time lags (time differences between any two epochs) of 2 years at 10.6 \micron. For one 
quasar, PG 1226+023, they measured structure functions in five bands, all showing $\sim0.1$ mag variations for a 
rest-frame time lag of 2 years.
Another example is the work of \cite{2006ApJ...639...46S}, who found a delayed response 
between the $K$- and $V$-band variability in four Seyfert 1 galaxies.
These authors showed that $K$-band light curves have a fairly tight time lag--luminosity relation, $\tau \approx L^{1/2}$, 
while broad line region (BLR) time lags are more scattered. The interpretation is that the $K$-band emission
is mainly coming from dusty clouds bounding the BLR, and that the
size scale of these dust clouds is very dependent on the AGN
luminosity, since they have to be colder than the dust sublimation temperature.

In this paper, for the first time, we investigate the variability of a truly large number of mid-IR sources, 
using the four-epoch {\it Spitzer} Deep Wide Field Survey (SDWFS; \citealt{2009ApJ...701..428A}) of the Bo{\"o}tes field of the NOAO Deep 
Wide Field Survey (NDWFS; \citealt{1999ASPC..191..111J}).  The two
goals are to characterize the mid-IR variability of a large sample of AGNs, and to use
the variability to identify additional AGNs.  While mid-IR quasar selection (\citealt{2005ApJ...631..163S}) works very well when  
the AGN dominates the luminosity (even for quasars behind dense stellar fields such as LMC/SMC; \citealt{2009ApJ...701..508K}), 
the method begins to fail as the luminosity becomes comparable to that
of the host (\citealt{2008ApJ...679.1040G,2010ApJ...713..970A}).   With mid-IR variability, we should be able to identify
such sources and extend the sample of mid-IR selected quasars to lower luminosities.  We also make
use of the extensive wavelength coverage of the field, including X-ray (XBo\"otes; \citealt{2005ApJS..161....1M}),
the earlier Infrared Array Camera (IRAC) Shallow Survey that became the first epoch of SDWFS \citep{2004ApJS..154..48E},
the 24 \micron~survey of the field (see \citealt{2005ApJ...622L.105H}),
and the FIRST \citep{1995ApJ...450..559B} and WSRT \citep{2002AJ....123.1784D} radio
surveys.  Based on these data, the AGN and Galaxy Evolution
Survey (AGES; C. S. Kochanek et al. 2010, in preparation) selected galaxies and AGNs at all these wavelengths for 
spectroscopic follow up using the Hectospec instrument on the MMT (\citealt{2005PASP..117.1411F}), providing roughly 23,500
redshifts in the field.  In Section~\ref{sec:data} we describe the data analysis, and in Section~\ref{sec:definitions}
we describe our variability selection criteria.  In Section~\ref{sec:results}, we present the results of this study.
The mid-IR structure function analysis is presented in Section~\ref{sec:sfunction}
followed by a summary in Section~\ref{sec:summary}.  Throughout this paper, we use a standard $\Lambda$CDM 
model with $(\Omega_{\Lambda}, \Omega_{\rm M}, \Omega_{k}) = (0.7, 0.3, 0.0)$ and $h=H_0/100=0.70$.  Where
needed, we use the galaxy and AGN templates of \cite{2010ApJ...713..970A}  to describe the mid-IR colors of 
various populations and to make any {\it K}-corrections or estimates of absolute luminosities.


\section{Data Analysis}
\label{sec:data}

The SDWFS is described in detail by \cite{2009ApJ...701..428A}. 
The images were taken with the IRAC instrument 
(\citealt{2004ApJS..154...10F}) on the \sst, covering wavelengths of 3.6, 4.5, 5.8, and 8.0 $\mu$m, hereafter referred to as the  
[3.6], [4.5], [5.8], and [8.0] bands, respectively. A single IRAC frame is $256\times256$ pixels. We work with full-field SDWFS mosaics 
constructed from the $\sim$20,000 IRAC images/epoch.   Each approximately $13600\times16700$ pixel mosaic covers 10.2 deg$^2$, 
centered at (R.A., decl.$)=($14:31:27, +34:10:43), with a pixel scale of 0.84 arcsec. Counting the original IRAC Shallow 
Survey (\citealt{2004ApJS..154..48E}) as epoch 1, the SDWFS data span a period of four years between 2004 and 2008, 
where the intervals between the epochs are approximately 3.5 years, 0.5 years, and 1 month 
(observations were done on 2004 January 10-14, 2007 August 8-13, 2008 February 2-6, and 2008 March 6-10). 
For each individual observation of a patch of the sky ($256\times256$ pixels), there is only a $\sim40$ s offset between 
the channel 1/3 and 2/4 observations. For a given SDWFS epoch, each patch of sky was observed $\sim3$ times over a 1--2 day period.
Each SDWFS epoch of the entire 10 deg$^2$ area took $\sim5$ days, and we treat it as a single image with a single Julian Date.
This does not affect our final conclusions, as we show later that there is little mid-IR AGN variability on these timescales.

\begin{figure}
\centering
\includegraphics[width=12cm]{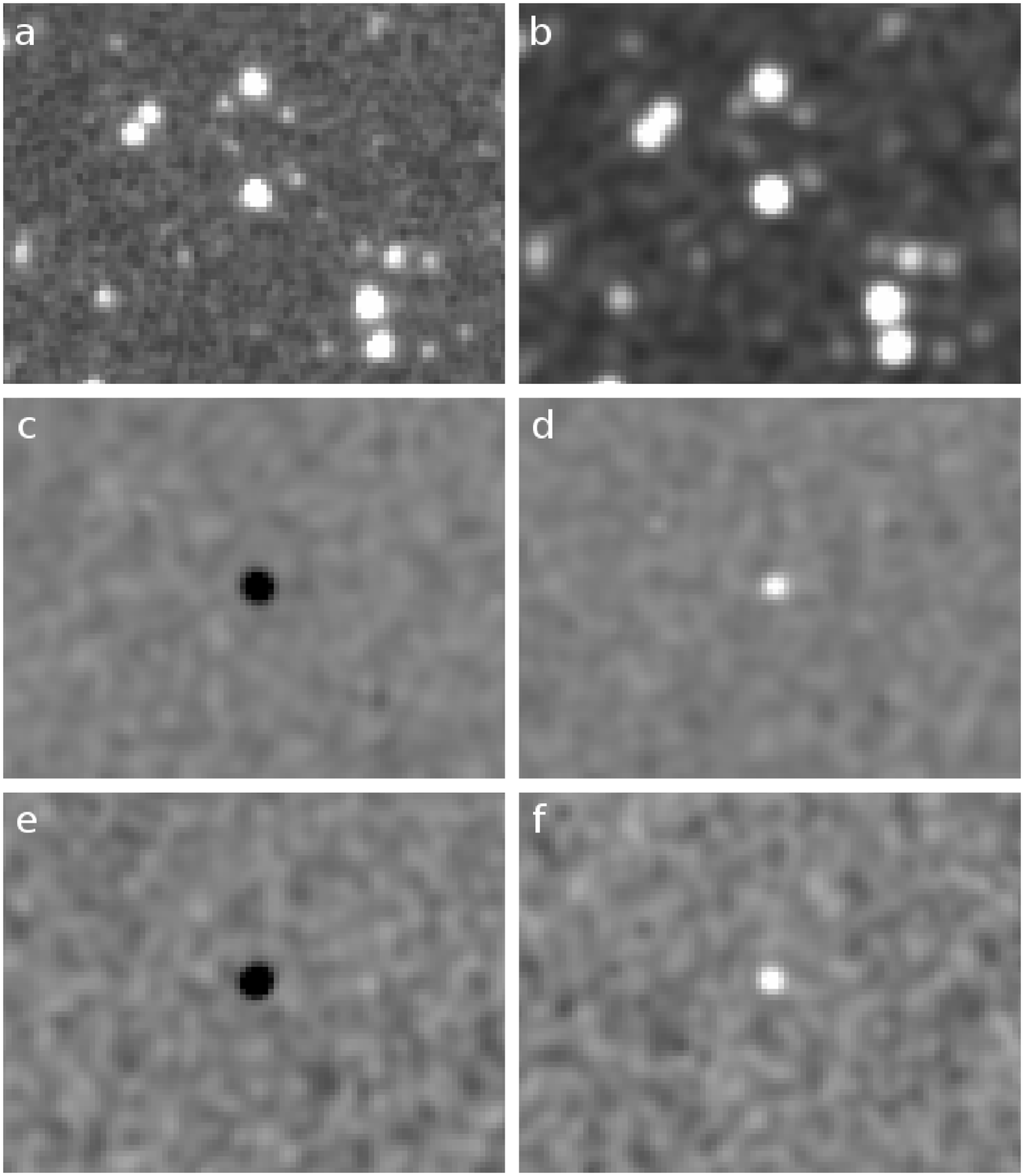}
\caption{Examples of data filtering and image subtraction. All six stamps cover the same area (1.2$\times$1 arcmin$^2$) and 
are centered on one of the most variable AGNs in the Bo{\"o}tes field, 
SDWFS~J143032.42+332158.8, with $z=0.43$, $\sigma_{12}=13.5$, and $r=1.0$ (see \S\ref{sec:definitions}).
Top row: stamp of the original 3.6 \micron~image (a). Before we do the image subtraction, 
we filter the images in order to convert the {\it Spitzer} IRAC's triangle-shaped PSFs into more Gaussian-like ones (b).
Middle row: subtracted images for the 3.6 \micron~channel. The AGN brightened 
by about 0.5 mag on a baseline of 4 years between epochs 1 (c) and 3 (d).
Black (white) color here reflects the source emitting less (more) flux than in the template image defined by the average of all four epochs. 
All non-variable objects are subtracted out.
Bottom row: subtracted images for the 4.5 \micron~channel. The AGN brightened by $\sim0.6$ mag between epochs 1 (e) and 3 (f), 
consistent with what is seen in the 3.6 \micron~channel.
\label{fig:image_analysis}}
\end{figure}

To study the variability of objects in the SDWFS field, we used the Difference Image Analysis (DIA) method 
(\citealt{1998ApJ...503..325A,2000AcA....50..421W}). While the variability itself is the 
subject of this paper, the DIA method also proved to be a useful diagnostic tool for constructing artifact-free 
mosaics (see \citealt{2009ApJ...701..428A}). The basic idea behind the DIA method is to match one image, called the template or 
reference image, both astrometrically and photometrically to another image of the same area of the sky.  Once they 
are astrometrically aligned, the template image (which is chosen to be of better quality) is convolved with a 
kernel function to match the point spread function (PSF) of the analyzed image. The two images are then subtracted 
from each other (Figure~\ref{fig:image_analysis}) to leave only variable objects, noise, and systematic problems (cosmic rays, artifacts, etc.). 
After some experimentation, we concluded that the  [5.8] and [8.0] maps were too shallow to perform DIA and obtain
reliable light curves, so our variability analysis is restricted to the [3.6] and [4.5] bands.

{\it Spitzer} PSFs have a ``triangular'' shape with extended diffraction-like patterns 
(more precisely they are ``multiplexer bleed'' or ``muxbleed,'' ``column pull-down,'' and 
``banding''\footnote{For details see the IRAC Data Handbook at \tt http://ssc.spitzer.caltech.edu/irac/dh/.}), and the PSF rotates between the epochs.
DIA works best on simple quasi-Gaussian PSFs, so we pre-filtered the images to have a simpler 
PSF structure (Figure~\ref{fig:image_analysis}). For each region, we 
scaled and median combined the brightest stars to build an empirical model of the PSF, $B$. We then linearly filtered the 
image $I$ to create a filtered image defined by the combination of Fourier transforms $\tilde{I}_F=\tilde{G}\tilde{I}/\tilde{B}$, 
that transforms the PSF from $B$ to $G$, where $G$ is a Gaussian broad enough to avoid noise amplification
(i.e., $\tilde{G}\to0$ faster than $\tilde{B}\to 0 $ for small scales/large wave numbers). 
The filtering process conserves fluxes and was carried out on a $3\times4$ grid of $4040\times4040$ pixel 
regions that overlap by 20 pixels along their edges. 
To save some computation time, each of these $4040\times4040$ regions was then cut into smaller $2040\times2040$ 
pixel sub-images (again each with 20 pixels of margin) for the 
DIA analysis. We combined all four epochs to build the template images and these templates were scaled and subtracted from each epoch using 
DIA (\citealt{2000AcA....50..421W}).  The brightest stars in the field produce artifacts due to subtraction residuals,
particularly from the ``diffraction'' spikes in the PSF.  
We therefore masked a region around all stars with $K<11$ mag in the 2MASS catalog of point sources (\citealt{2003tmc..book.....C}). 
The masking radius was $260(1-K/11)$ arcsec, which is large enough to include the spikes. We used the Two Micron All Sky Survey (2MASS) catalogs because 
these bright stars were saturated and not included in the SDWFS catalogs. The masked regions constitute $\sim3$\% of the area.

The [3.6], [4.5], [5.8], and [8.0] Vega magnitudes, and the coordinates (J2000.0) are taken from the SDWFS catalogs (\citealt{2009ApJ...701..428A}). 
Throughout this paper, we use 4 arcsec diameter aperture magnitudes with aperture corrections to 24 arcsec. 
The original 3.6 \micron~full 
SDWFS catalog contains 677,522 objects (catalog DR1.1), but we cropped it to
515,055 objects (hereafter called the ``cropped sample'') as variations in coverage between epochs 
reduced the usable area for variability analysis by about 20\%, to 8.1 deg$^2$. Using DIA's PSF fitting routines, we could 
construct light curves (at least three epochs) for 492,420 objects simultaneously in both [3.6] and [4.5]. The final number of objects
used in this analysis after masking regions around bright stars is 474,179. 
While we include the results for all objects from 
the cropped sample in Table~\ref{tab:SDWFSvar}, 
in the analysis that follows we consider only the sources with at least three epochs. 
Table~\ref{tab:SDWFSlc} provides the DIA light curves for all SDWFS sources from the cropped sample.

In order to understand the nature of the variable sources, we matched the SDWFS sources to several of the available data sets
for the Bo\"otes field, focusing on information related to AGN activity.  Thus, we matched to the X-ray and radio catalogs,
to the subset of 24 \micron~sources with high probabilities of being AGNs, and to the AGES spectroscopic data.
We used the X-ray catalogs from the XBo{\"o}tes \chandra~survey covering 9.3 deg$^2$ of the Bo{\"o}tes field in the
0.5-7 keV band (\citealt{2005ApJS..161....1M, 2005ApJS..161....9K,2006ApJ...641..140B, 2008ApJ...679.1040G}). We matched
a version of the \cite{2005ApJS..161....1M} catalogs extended downward in flux from $3293$ sources with 4 or more counts
to $4642$ sources with 2 or more counts.  While using 2 count (about $4 \times 10^{-15}$ erg cm$^{-2}$ s$^{-1}$)
X-ray sources sounds peculiar, the backgrounds in XBo\"otes are so low that almost all such sources with mid-IR identifications 
will be real, particularly if the match is to one of the relatively rare mid-IR variable sources.  Using the Bayesian
matching method of  \cite{2006ApJ...641..140B}, we identified $3363$ X-ray sources with at least two epochs of SDWFS data.  
We used the same procedures to match to the 20 cm FIRST \citep{1995ApJ...450..559B} and 1.4 GHz WSRT  (\citealt{2002AJ....123.1784D})
radio catalogs.  Restricting the matching to likely radio point sources in the FIRST and WSRT surveys, we found $322$ and $2118$ matches,
respectively, to SDWFS sources with at least two epochs of data.  We know from AGES that most 24 \micron, optical point sources satisfying 
the criterion that $I> 18-2.5\log(F_{24}/\hbox{mJy})$ are quasars (\citealt{2006ApJ...638...88B, 2010ApJ...708..584E,2010arXiv1001.4529A}), 
where the optical criterion is used to eliminate stars.  Finally, we matched to the AGES redshift survey, finding 
$19,885$ ($18,288$, $1387$, $497$) SDWFS sources (with redshifts $z>0, 1, 2$).  The
modified SDSS pipeline used for the AGES redshifts (HSRED; \citealt{2008ApJ...682..919C}) provides an estimate of the
best fit spectral template used for the fits that can be used to classify objects.  More generally, we have all the
galaxy and AGN target selection codes used by AGES to select and prioritize spectroscopic targets. 


\section{Variability}
\label{sec:definitions}

For each light curve $i$ (from Table~\ref{tab:SDWFSlc}) with at least three epochs in both the [3.6] and [4.5] bands, we computed the 
standard deviation $v([X])^i$ for that light curve as a function of (Vega) apparent magnitude $m([X])^i$, 
as shown in Figure~\ref{fig:var3645}. We also calculated the variability covariance $C_{12}^i$ and Pearson's 
correlation $r^i$ coefficients between the two channels, where
\begin{eqnarray}
v([X])^i & = & \left({\frac{1}{N_{\rm epochs}-1}\sum_{j=1}^{N_{\rm epochs}}\left(m([X])^i_j-\langle m([X])^i \rangle\right)^2}\right)^{1/2},\\
C_{12}^i & = & \frac{1}{N_{\rm epochs}-1}\sum_{j=1}^{N_{\rm epochs}}\left(m([3.6])^i_j-\langle m([3.6])^i \rangle\right)\left(m([4.5])^i_j-\langle m([4.5])^i \rangle\right), \\
r^i & = & \frac{C_{12}^i}{v([3.6])^i v([4.5])^i},
\end{eqnarray}
and $\langle m([X])^i \rangle$ is the average magnitude for the $i$th light curve in channel $X$. 
The correlation coefficient is limited to the range $-1 \leq r \leq 1$, 
where $r=1$ ($-1$) means that the [3.6] and [4.5] light curves are perfectly correlated (anti-correlated).  The distribution of $v([X])^i$ is dominated by 
contributions from noise (and any systematic errors) rather than variability, so we need to define a threshold
for selecting candidate variables from these distributions.  Let $v_m([X])$ be the median of the $v([X])^i$ as a function
of magnitude and $\sigma([X])$ be the standard deviation of the $v([X])^i$ values around this median. Sources are more likely
to be true variables as their distance, $(v([X])^i-v_m([X]))/\sigma([X])>0$, from the median increases, and if the [3.6]
and [4.5] variabilities are strongly correlated. A reasonable metric for the joint significance of the variances in the light curves is 
\begin{equation}
\sigma_{12}^i=\left(
     \left({v([3.6])^i-v_m([3.6]) \over \sigma([3.6])}\right)^2
    +\left({v([4.5])^i-v_m([4.5]) \over \sigma([4.5])}\right)^2
     \right)^{1/2},
\end{equation}
which quantifies the degree to which the object deviates from the median variances in both bands. Objects with
positive excess variances (large $\sigma_{12}$ and $v>v_m$) and strong correlations $r\simeq 1$ between the two bands are the 
best candidates for true variables. 

Figure~\ref{fig:covariance} shows the distribution of correlation strengths 
$r$ for sources with $\sigma_{12}$ exceeding a range of thresholds.  We see that there is a peak for highly 
correlated sources $r \simeq 1$ whose strength increases as the variability significance $\sigma_{12}$ 
increases.  This demonstrates that the variability in the two bands is tightly correlated for real variables
and that we are likely to produce a relatively clean sample of variables (the false positive rate is 6\%--7\% for AGNs in the Photometry Group 3 defined below) 
if we restrict ourselves to $ r > 0.8$ and $\sigma_{12} > 2$.   
Variability selection is always a tradeoff between completeness and contamination, so we will consider
sources in three different variability levels, all restricted to have correlations $r>0.8$ between
the bands. To avoid confusion, the Variability Levels correspond directly to the values of $\sigma_{12}$.
Variability Levels 2, 3 and 4 correspond to sources with excess variances and 
significances exceeding $\sigma_{12} >2$, $3$ and $4$, respectively, with Variability Level 2 being the
most complete and the most contaminated, and Variability Level 4 being the least complete and least 
contaminated. 

We estimate the fraction of false positives by assuming that the density
of uncorrelated sources ($r<0.5$) in Figure~\ref{fig:covariance} represents the density of
false positives amidst the highly correlated ($r>0.8$) sources. If the number of identified variables is
$N_{r>0.8}$, that is sources with $\sigma_{12} > 2$ (or 3, 4) and $r>0.8$, and the number of sources with 
$\sigma_{12} > 2$ (or 3, 4) and $r<0.5$ is $N_{r<0.5}$, then we estimate that the number of false positives is
$F=(0.2/1.5)N_{r<0.5}$, where $0.2/1.5$ is the ratio of the intervals $r>0.8$ and $-1 \le r \le 0.5$.
The fraction of the identified variables that are false positives is then $F/N_{r>0.8}$.

Our ability to characterize sources also depends on the magnitudes of the object, because this determines
how well we can measure the mid-IR colors.  We divide sources into three photometry groups based on their mid-IR magnitudes.
\begin{itemize}  
\item Photometry Group 1 consists of all objects that after masking have three or four epochs of data, with no constraints on either 
    their magnitudes or magnitude uncertainties.  There are 474,179 Photometry Group 1 objects after masking, of which 
    5107, 2208, and 1071 meet the Variability Level 2, 3, and 4 criteria.

\item Photometry Group 2 is the subset of the Photometry Group 1 sources for which we can measure the [3.6]$-$[4.5] color.  These
   sources must satisfy $[3.6] < 19.7$ mag and $[4.5] < 19.3$ mag with uncertainties in these
   bands smaller than 0.1~mag.  There are 213,594 Photometry Group 2 objects, of which 1557, 668, and 359 meet
   the Variability Level 2, 3, and 4 criteria.

\item Photometry Group 3 is the subset of Photometry Group 2 objects for which we can also measure the [5.8]$-$[8.0] color.  These
   sources are required to have uncertainties smaller than $<0.2$ mag for the 5.8 and 8.0 \micron~bands.  There
   are 39,522 Photometry Group 3 objects, of which 775, 428, and 260 meet the Variability Level 2, 3, and 4 criteria.
\end{itemize}
Table~\ref{tab:SDWFSvar} provides these significance statistics for the SDWFS sources. 
Note, however, that our catalogs are limited to the objects detected on the template image -- any transient which does not produce 
a detectable source in the template image, which is just the average of four epochs, will not be identified.
The most variable, highly correlated ($r>0.8$) sources in the field are AGNs, peaking at $\sigma_{12}\approx 20$.

We will make extensive use of the [3.6]$-$[4.5] versus [5.8]$-$[8.0] mid-IR color-color distribution 
to characterize the sources. Figure~\ref{fig:colcolintro} shows the distribution of all Photometry Group 3 SDWFS sources in this space.   
\cite{2004ApJS..154..48E} suggested that for bright sources the red plume in [3.6]$-$[4.5] color was likely to 
be dominated by AGNs.  This was confirmed by \cite{2005ApJ...631..163S} based on the first year of 
spectroscopic observations by the AGES survey. \cite{2005ApJ...631..163S} defined a selection region,
the ``AGN wedge'', for mid-IR AGN selection that is also shown in Figure~\ref{fig:colcolintro}.  For these Vega colors,
stars lie at color 0 unless they have dusty winds or disks to produce a mid-IR excess.  Using the
spectral energy distribution (SED) models of \cite{2010ApJ...713..970A}, which are derived by fitting the
SEDs of Bo\"otes sources, we can illustrate the locations of various extragalactic populations
as a function of redshift.  Early-type galaxies at low redshifts have colors similar to stars and
then become redder in [3.6]$-$[4.5] for $z>1$ at roughly fixed [5.8]$-$[8.0], tracking along the
left edge of the AGN wedge. They become red in [3.6]$-$[4.5] due to the flux ``peak'' near rest-frame wavelength of 
1.6 \micron, which passes through the 
4.5 \micron~band at a redshift of 1.5--2.1. Late-type galaxies at low redshift have red [5.8]$-$[8.0] colors at low redshift 
due to the polycyclic aromatic hydrocarbon (PAH) emission in the 8 \micron~band, then move horizontally to bluer [5.8]$-$[8.0]
along the bottom of the AGN wedge as the feature is redshifted out of the 8 \micron~band,
and finally follow the early-type galaxies up along the left edge of the selection region.  Pure
AGNs remain inside the AGN wedge except near $z\sim 4.5$ where they make an excursion out of
the AGN wedge when the H$\alpha$ line lies in the 3.6 \micron~band.  If we use an equal
combination of an AGN and a host galaxy, defined by the total emission from 0.1 to 30 \micron, 
the combined SED is somewhat bluer.  With further increases in the host luminosity relative
to the AGN, the source drops below the bottom of the AGN wedge.  

In fact, the distribution of 
the IRAC colors for the XBo\"otes X-ray sources does extend below the \cite{2005ApJ...631..163S} AGN wedge
\citep{2008ApJ...679.1040G}.  Based on \cite{2008ApJ...679.1040G}, we define a ``modified AGN
wedge'' that encompasses most X-ray sources. 
This modified AGN wedge is defined by $[3.6]-[4.5]<2.5\times([5.8]-[8.0])-0.6$ mag (left side), 
$[3.6]-[4.5]>2.5\times([5.8]-[8.0])-3.5$ mag (right side), and $[3.6]-[4.5]>0.2\times([5.8]-[8.0])+0.08$ mag (bottom).  
Because we are identifying AGN candidates based on
their mid-IR variability, we should be able to identify them in this $[3.6]-[4.5]$ blueward extension of the 
\cite{2005ApJ...631..163S} selection region despite any host contamination. The caveat, however, is that this blueward extension
is dominated by false positives (galaxies) if used to select AGNs based solely on their mid-IR colors without
additional information such as variability or X-ray emission.
We also define a ``stellar'' box which should contain most normal stars. The edges of this box are defined by 
$-0.15$ mag $<[3.6]-[4.5]<0.15$ mag and $-0.15$ mag $<[5.8]-[8.0]<0.15$ mag.
The coldest stars, reaching into the brown dwarf regime, become
red in $[3.6]-[4.5]$ due to methane absorption in the 3.6 \micron~band,
but have $[5.8]-[8.0]$ colors near zero and thus stay to the left of
the AGN wedge (\citealt{2007ApJ...663..677S,2010AJ....139.2455E}).  

Beyond these mid-IR color diagnostics,
we also use the extensive multi-wavelength and spectroscopic data for the field to characterize the sources. 
We know that the variable sources must be dominated by AGNs and a small fraction
of stars, so the basic check on our selection criteria is that our selection limits on $\sigma_{12}$ and $r$ 
preferentially select sources with AGN signatures, without introducing large numbers of contaminating sources.


\section{Results}
\label{sec:results}

We will now characterize the variable sources using three approaches. First, we simply examine the distribution
of objects in mid-IR color.  Second, we examine the variability of AGN candidates identified at other wavelengths.
Finally, we consider the spectral properties of the objects with known redshifts.  

Figure~\ref{fig:colcol_hist} and Table~\ref{tab:resultsSDWFS} characterize the distribution of objects in mid-IR color.  
The Photometry Group 3, Variability Level 2 candidates are
concentrated in either the modified AGN wedge, with a distribution very similar 
to that of X-ray sources (\citealt{2008ApJ...679.1040G}; see Figure~\ref{fig:colcolintro}),
or in the stellar box. While only 0.5\% of the Photometry Group 3 objects outside these two regions are Variability Level 2 
sources, 7.4\%, 4.6\%, and 1.1\% of the sources in the old AGN, modified AGN, and stellar regions are Variability Level 2 sources.
The estimated false positive rates 
are 44\%, 6\%, 7\%, and 58\% for these four regions (outside, old AGN wedge, modified AGN 
wedge, stellar box) in the color-color space, respectively. This simply means that, for example,
of the 7.4\% variable sources in the old AGN wedge 6\% are false positives, so in fact
7.4\%$\times(1-0.06)=7.0$\% of all objects are real variables and 7.4\%$\times0.06=0.4$\% of all objects are falsely identified as variable.
Increasing the variability significance level reduces the variable fractions
roughly in the same proportion, suggesting that even the Variability Level 2 candidates are dominated by true
variables. The distribution also validates our assumption that the variability is dominated by
AGN -- 75\% of the candidates lie in the modified AGN region for Variability Level 4. 
The selection remains pure even for Variability Level 2 because the requirement that the [3.6]
and [4.5] light curves are tightly correlated ($r>0.8$) eliminates most of the contamination even
when we only require $\sigma_{12}>2$ (see Figure~\ref{fig:covariance}).
Most variable sources in the modified AGN wedge are known to be AGNs based on other evidence.
Only 16\%, 10\%, and 7\% of the Photometry Group 3, Variability Level 2, 3, and 4 sources lack
photometric or spectroscopic signs of AGN activity. Most of these, however, are likely to be false 
positives given our estimated false positive rates of 7\%, 4\%, and 3\%.
For the Photometry Group 2 objects, where we lack accurate 
[5.8]$-$[8.0] color, we can only investigate the distribution in [3.6]$-$[4.5] color, and the
side panel to Figure~\ref{fig:colcol_hist} shows the fraction  of variable sources as a function of this color.  
Not surprisingly, the fraction is high ($\sim10$\%) when the color is red, corresponding to AGNs, and then drops dramatically
for colors consistent with normal galaxies.  The fraction of variables in the other two photometry groups
are dramatically lower than for Photometry Group 3 for two reasons. First, a source must be intrinsically
more variable in order to meet any variability criterion because the intrinsic measurement
errors are larger.  Second, with less well-measured colors, objects are scattered from their
true location in the color-color space, and the differential scattering of galaxies into the AGN regions and 
AGNs out of these regions dilutes the selection criterion. 

The 86 (Photometry Group 3, Variability Level 2) variable objects inside the stellar box are red, with optical 
colors of $R-I\gtrsim1$ (i.e., $V-I\gtrsim2$), so they must be dominated by 
red giants or main-sequence stars. The red and asymptotic giant branch stars include the OGLE Small Amplitude 
Red Giants (OSARGs;  e.g., \citealt{2004AcA....54..129S}) with $I$-band amplitudes of $\sim0.1$ mag and periods
between 10 and 1000 days, the Long Period Variables (LPVs; e.g., \citealt{2009AcA....59..239S}) with amplitudes 
of up to $\sim3$ mag and periods 10-1000 days, and the Long Secondary Period variables (LSPs; e.g., \citealt{2007ApJ...660.1486S}) 
with periods 30--1000 days. However, given their absolute magnitudes of $M_I\approx-4$ mag, they would have 
to be at a distance of $\sim200$ kpc to match the observed $R$, $I$, and IRAC magnitudes.
Since this scenario seems unlikely, the variable objects inside the stellar box could be the main-sequence flaring 
M dwarfs (e.g., \citealt{2009AJ....138..633K}) with absolute magnitudes of $M_I\approx7$ mag, and distances of 
up to 1 kpc. Based on \cite{2009AJ....138..633K}, we expect to find $\sim10$ M dwarf flares in our data. The $R-I$ and 
$I-[3.6]$ colors of the variable objects inside the stellar box are inconsistent with the galaxy templates from \cite{2010ApJ...713..970A}.
We have not pursued these sources in any detail because they are almost certainly quotidian variable stars and false positives.
The estimated false positive rate of 58\% is very high.

Next, we examine the variability of sources whose X-ray, MIPS 24 \micron~or radio fluxes indicate
that the source is an AGN.  The color distributions of these sources are shown in Figure~\ref{fig:col-col_QSO} and the
variability statistics are summarized in Table~\ref{tab:resultsAGES}.  Here we focus on Photometry Group 1 since we are
not dependent on having mid-IR colors to interpret the results.  First, we see that a 
significant fraction of the X-ray sources show mid-IR variability, with 12\% ($=12.1\%\times(1-0.04)$ accounting for false positives) being variable
even with no limits on the mid-IR photometric uncertainties.  The higher fractions in Photometry Groups
2 and 3 are largely due to our improved ability to detect a given level of variability as we
require smaller photometric uncertainties.  The MIPS quasar candidates, which are found to
be either AGNs ($\sim 80\%$) or optically unresolved star forming galaxies upon spectroscopic
examination, also show high variability rates, with 11\% ($=11.7\%\times(1-0.05)$) of the Photometry Group 1 objects showing
variability at Variability Level 2.  The variable fraction again increases as we go to the brighter
photometric groups.  The two sets of radio sources show significantly lower levels of variability.
This is consistent with the results of \cite{2009ApJ...696..891H} and \cite{GS2010} 
who show that, as compared to X-ray and mid-IR selected AGNs, radio-selected AGNs have lower 
Eddington ratios and are much less likely to be AGN dominated.  Thus, variability in radio-selected 
AGNs will be more challenging to discern relative to the photometric noise due to the host galaxy.

Finally, we compare the variable sources with the template types used to estimate the
source redshifts in the AGES survey, as shown in Figure~\ref{fig:col-col_SDSS} and 
summarized in Table~\ref{tab:resultsSpec}.  
This is somewhat redundant with the previous two 
tests because the objects targeted as AGNs are chosen to be in the IRAC AGN wedge,
to show X-ray or radio emission, or to be optical point sources with non-stellar 24 \micron~fluxes.  
The success rate for the spectroscopic identification of candidate AGNs in
AGES is quite high.  The sample of stars with spectra is very heterogeneous. Roughly
half of the stars are F and G stars observed by AGES as spectrophotometric 
calibrators.  Most of the remainder are K and M stars which were flagged as
candidate AGNs either because of their X-ray fluxes or the presence of a mid-IR
excess.  We again see that the fraction of variable sources is highest for 
the AGNs, with roughly 11\% meeting the Level 2 variability criterion.

In summary, small relatively shallow mid-IR variability surveys like SDWFS are a workable but not a highly efficient means
of identifying AGNs. Roughly speaking, the efficiencies are of order 10\% for variability selection criterion
providing relatively low false positive rates. It does, however, overcome the weaknesses of mid-IR color selection
to identify AGNs outside the \cite{2005ApJ...631..163S} selection region. With difference imaging, the host galaxy flux
that moves the source outside this color region ceases to be a problem other than through its effects on the photometric noise.

\section{Mid-IR Structure Functions}
\label{sec:sfunction}

The structure function is a measure of the expected changes in an object's flux as a function of the time between the measurements.
The time lag $\tau $ is defined as the time difference between any two data points in a light curve, 
where the observed time lag is $\tau = |t_i - t_j|$ and the rest-frame time lag is $\tau = |t_i-t_j|/(1+z)$.
For each time lag, one wants to find all pairs of points in a light curve and calculate the variance of the magnitude differences, $\Delta m = m_i-m_j$.
Such a variance can also be calculated by averaging over many objects of the same class (i.e., AGNs) for each time lag 
between any two epochs, as we do in this paper.

Given the depth of the AGES redshift survey, all $z>1$ objects are spectroscopically confirmed AGNs (see Figure 2 in \citealt{2005ApJ...631..163S}).
We measured the mid-IR structure function of $z>1$ Photometry Group 3 AGNs to fixed magnitude
limits of $[3.6]<16$, $17$, and $18$ mag.  Only $\sim 10\%$ of these objects were
actually detected as significantly variable sources (see Table~\ref{tab:resultsSpec}), so most of the observed
variance in the magnitudes is at the noise level.  This is a rather different regime from
most optical studies, where the variance is dominated by variability, and makes
it critically important to correctly characterize the contribution to the
variance from the measurement errors.  We do this by matching each AGN with
four unresolved galaxies of nearly identical observed flux.  We chose these galaxies
to be Photometry Group 2 and 3 objects with colors $[3.6]-[4.5]<0.1$ mag and 0.5 mag $<[5.8]-[8.0]<1.5$ mag, thus avoiding AGNs.  
We then compute the structure function as
\begin{equation}
   S(\tau) = \left[
     { 1 \over N_{\rm qso}(\tau) } \sum_{i<j} \left( m(t_i) - m(t_j) \right)^2_{\rm qso} -
     { 1 \over N_{\rm gal}(\tau) } \sum_{i<j} \left( m(t_i) - m(t_j) \right)^2_{\rm gal}\right]^{1/2}.
\label{eqn:sf}
\end{equation}

In addition to the simple calculation over the $160$, $690$, and $1100$ AGNs
with [3.6]$<16$, $17$, and $18$~mag, we also calculated the structure functions for
1000 bootstrap resamplings of both the AGN and galaxy lists.  These give consistent
results, and we generally report the median result of the bootstrap samples and the
region encompassing 68.3\% of the samples about the median.  The small values at short lags
are very sensitive to any systematic problems in the noise estimate, and while this
empirical approach to correctly characterizing the noise appears to work well, our
estimates of the amplitude of the structure function on the longer timescales are
probably more robust than our estimate of the slope of the structure function.

Figure~\ref{fig:sf_obs} shows the resulting observed and rest-frame structure functions for the
[3.6] and [4.5] bands as a function of magnitude limit.  We generally see more
variability on longer timescales.  We fit our structure functions with a power-law
\begin{equation}
     S(\tau) = S_0 \left( { \tau \over \tau_0 } \right)^\gamma
\end{equation}
using $\tau_0=4$~years ($2$~years) for the observed (rest-frame)
estimates.  Table \ref{tab:sf_fits} presents the resulting fits.  The [3.6] and [4.5]
structure functions are consistent with each other, but also significantly steeper than
the $i$-band structure function from an ensemble of SDSS quasars (\citealt{2004ApJ...601..692V}), with $\gamma = 0.303 \pm 0.035$, and higher
amplitude, with $S_0=0.20$~mag for $\tau_0=2$~years.

The steeper mid-IR slope could be explained by the emission being dominated
by larger physical scales than the optical emission.  At these redshifts,
the observed frame mid-IR corresponds to the rest-frame near-IR, where
the emission has contributions from both the disk and hot dust.  If
we focus on the more compact disk emission, the characteristic scale of an
accretion disk for emission at wavelength $\lambda$, defined by the
point where the disk temperature matches the photon energy, is
\begin{equation}
      R_\lambda \simeq 10^{16} \left( { \lambda_{\rm rest} \over \mu \hbox{m} }\right)^{4/3}
         \left( { M_{\rm BH} \over 10^9 M_\odot } \right)^{2/3}
         \left( { L \over \eta L_E }\right)^{1/3} \rm cm,
\end{equation}
where $L/L_E$ is the Eddington ratio and $L= \eta \dot{M} c^2$ defines the
radiative efficiency $\eta$.  Taking $L/L_E=1/3$, $\eta=0.1$, and $z=1$, the optical
emission for $M_{\rm BH} = 10^9 M_\odot$ comes from a region only a light day
($0.003$~light years) in size, while the mid-IR scale is approximately 10 light days
($0.03$~light years).   Thus, our shortest lags correspond to the fastest possible
response time given the size of the emission region since the typical
AGES quasar has $M_{\rm BH} \simeq 10^{8.5} M_\odot$ (\citealt{2006ApJ...648..128K}).
The orbital, thermal and viscous timescales are even longer, and Kelly
et al. (2009, also see \citealt{2010ApJ...708..927K}) find that the optical
variability can be characterized by an amplitude and a timescale that
might be associated with thermal timescales.  In any case, the mid-IR
emission is coming from scales so large that it is hard to respond to
any driving process on the timescale of our shortest observed lags.  This should
suppress variability on these short timescales compared to the optical
emission, which can respond more easily, leading to a steeper mid-IR structure
function.

Next we subdivided the sample into two redshift bins, $1<z<2$ and
$2<z<3$, and then divided the objects in these bins at the median
value of their absolute magnitudes.   Figure~\ref{fig:sf_bin} shows the structure
functions of these subdivided samples as compared to the best fits
to the full samples.  At least for the well-determined values at
longer lags, there is a pattern  that shorter
rest-frame wavelengths vary more than longer rest-frame wavelengths. Also for $1<z<2$ AGNs,
the lower luminosity systems vary more than the higher luminosity systems.
This is consistent with earlier findings in the optical part of the spectrum  
(e.g., \citealt{1994MNRAS.268..305H,2002MNRAS.329...76H,2004ApJ...601..692V,2005AJ....129..615D}). In general,
the trend of more variability with longer lags also holds,
although there are some exceptions in the [4.5] results.

We also subdivided the data into eight smaller redshift--absolute magnitude bins and performed a global fit that includes a dependence 
on wavelength in the following form (see also \citealt{2004IAUS..222..525I})
\begin{equation}
S(\tau) = A\left[1+B\left(\langle M_J \rangle+25\right)\right]\left(\frac{\tau_{\rm rf}}{\rm 2~years}\right)^C\left(\frac{\langle \lambda \rangle}{1~\micron}\right)^{-D},
\end{equation}
where $A,B,C$, and $D$ are parameters of the fit, $\langle M_J \rangle$ is the average absolute $J$-band 
magnitude in a bin (the rest-frame $M_J$ magnitudes are taken from \cite{2010arXiv1001.4529A}), $\tau_{\rm rf}$ is the rest-frame time lag, 
and $\langle \lambda \rangle$ is the average rest-frame wavelength in each bin. 
We either fit (1) all the data or (2) exclude the short time-lags ($\tau<0.1$ year) as the structure 
function contains little or no signal in this regime. For case (1) we found
$A(1) =  0.13 \pm 0.01$,
$B(1) =  0.19 \pm 0.03$,
$C(1) =  0.41 \pm 0.05$,
$D(1) = -0.04 \pm 0.12$, while for case (2) we found
$A(2) =  0.13 \pm 0.01$,
$B(2) =  0.19 \pm 0.03$,
$C(2) =  0.46 \pm 0.06$,
$D(2) = -0.02 \pm 0.13$.
We find no detectable dependence of the structure function on wavelength, even though the data cover the redshift range of $z=1$--$3$, 
the absolute magnitude range of $M_J=-24.5$ to $-26.5$ mag, and the rest-frame wavelengths $\lambda_{\rm rf}=1$--$2$ \micron. 
As compared to optical studies (e.g., \citealt{2010arXiv1004.0276M}), the mid-IR structure function appears to have stronger dependence on the 
luminosity (parameter $B$), and no dependence on wavelength (parameter $D$). Note that the large uncertainty in $D$ ($\sigma_D=0.13$) 
can potentially hide, within the 2--$3\sigma$, the wavelength dependence of $D\simeq0.3$ observed in optical studies. Another possibility is 
that the mid-IR wavelength dependence is much weaker than that observed in optical.

Table~\ref{tab:sf_fits1} shows the variability amplitudes at the observed time lag of 4 years  
for various classes of objects from Figure~\ref{fig:colcolintro} assuming a fixed slope of $\tau^{1/2}$. Objects with $z>1$ are AGNs, and
show uniform and consistent variability amplitudes of $\sim0.09$ mag as a function of magnitude in both channels for fixed $\gamma=0.5$.
The X-ray selected sources inside the modified AGN wedge are expected to be AGNs, and not surprisingly,
they also show similar variability amplitudes of $\sim0.10$ mag to the $z>1$ objects.
The mid-IR counterparts of the 24 \micron~(MIPS) selected
sources inside the modified AGN wedge, again, have almost identical variability amplitudes. 
The radio selected objects, however, have a drastically
different distribution in their mid-IR colors (top right panel in Figure~\ref{fig:colcolintro})
from that of AGNs, suggesting that a large fraction of them may not be AGNs. The structure functions 
based on the radio-selected objects inside the modified AGN wedge may be driven by non-AGN objects, and we 
find somewhat smaller variability amplitudes of $S_0=0.05\pm0.01$ mag and
 $S_0=0.06\pm0.02$ mag in [3.6] and [4.5], respectively.

\section{Increasing Completeness with Future Surveys}

\begin{figure}
\centering
\includegraphics[width=12cm]{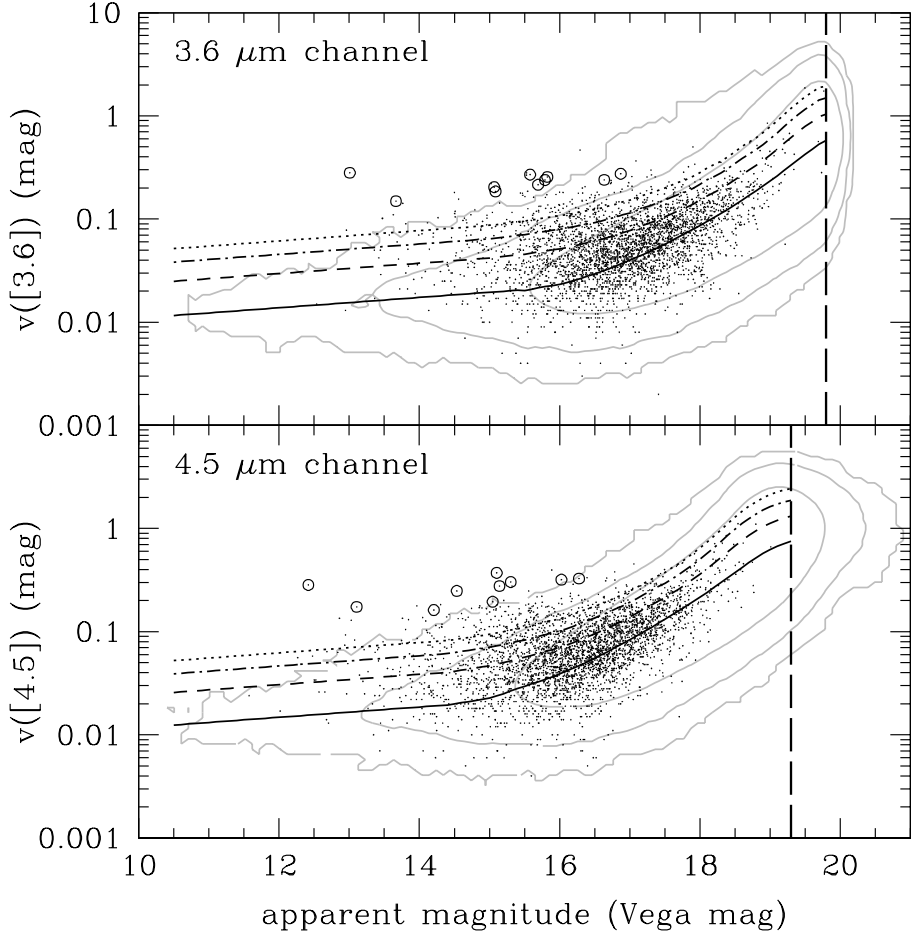}
\caption{Variability in the 3.6 \micron~(top) and 4.5 \micron~(bottom) channels as a function of magnitude for 
all objects with four epochs of data shown as contours. The objects are binned into 0.1 mag and 0.05 log($v$) bins. 
The contours are drawn for 2, 10, and 100 objects per bin, counting from the outer contour. The lower solid line 
shows the median $v_m$ calculated in 0.1 mag bins, and the upper three dashed, dash-dotted, and dotted lines show 
variability significances $(v-v_m)/\sigma=1$, 2, and 3  (from bottom to top) with respect to this median. We limit our Photometry Group 2 
and 3 samples to objects with $[3.6] < 19.7$ mag, and $[4.5] < 19.3$ mag (vertical dashed lines). 
Dots mark $\sim3200$ Photometry Group 3 AGNs from the modified AGN wedge, selected as either X-ray or 24 \micron~AGN candidates 
or having a QSO spectroscopic template code. Note that they tend to have higher average variability than the typical sources,
but with amplitudes easily masked by the photometric uncertainties. Open circles mark objects with $\sigma_{12}>10$ and $r>0.8$. 
\label{fig:var3645}}
\end{figure}

\begin{figure}
\centering
\includegraphics[width=16cm]{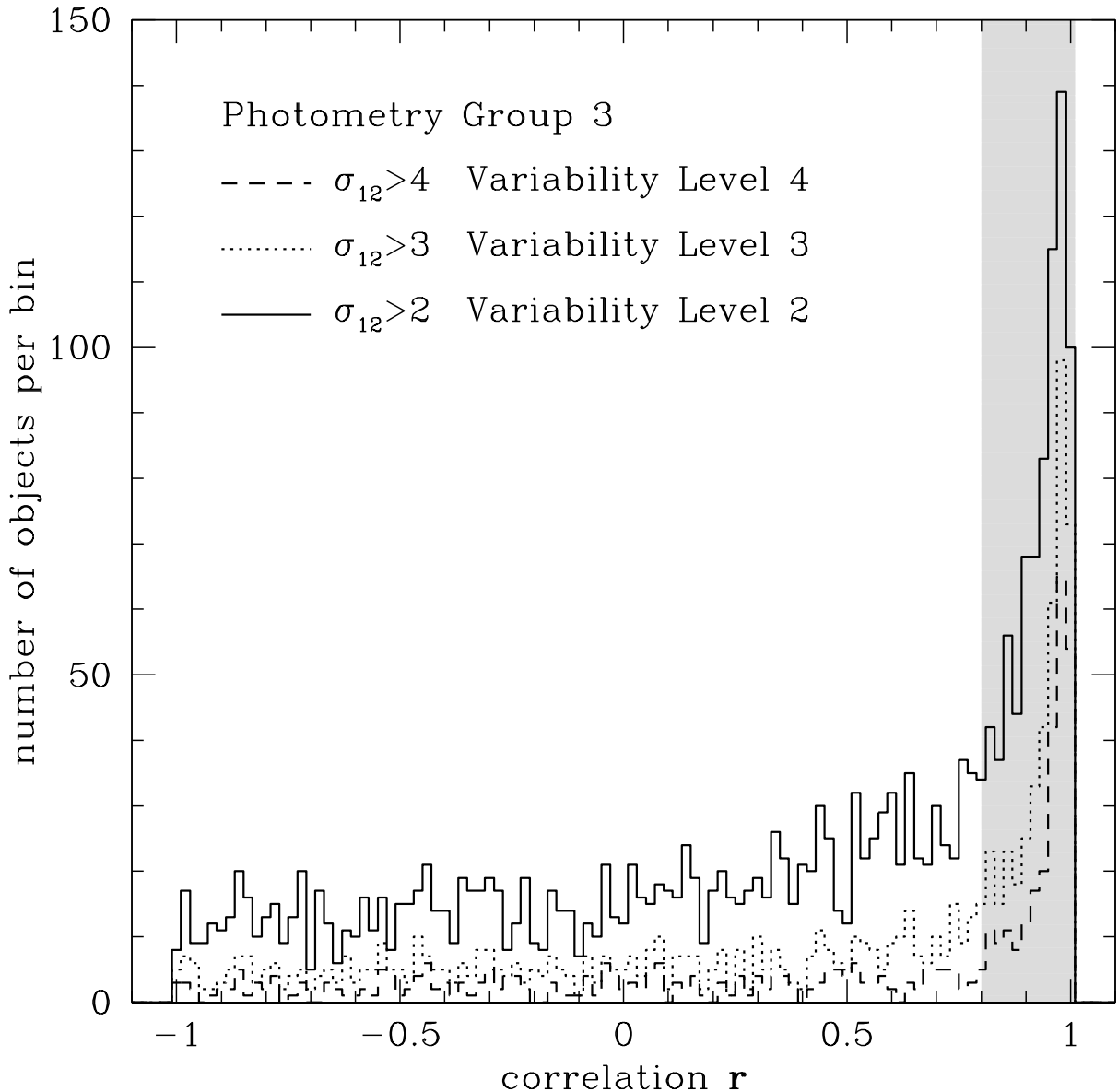}
\caption{Correlation between the 3.6 \micron~and 4.5 \micron~light curves for sources with measured $[5.8]-[8.0]$ colors
(``Photometry Group 3'' sources; see Section \ref{sec:definitions}) with four epochs of data for
Variability Level 2, 3, and 4 ($\sigma_{12}>2$, $>3$, and $>4$) sources.  We only consider sources in the shaded, 
highly correlated $r>0.8$ region as true variables, and we use the density of the poorly correlated $r<0.5$ 
sources to estimate our false positive rates (Tables~\ref{tab:resultsSDWFS}--\ref{tab:resultsSpec}).
\label{fig:covariance}}
\end{figure}

\begin{figure}
\centering
\includegraphics[width=12cm]{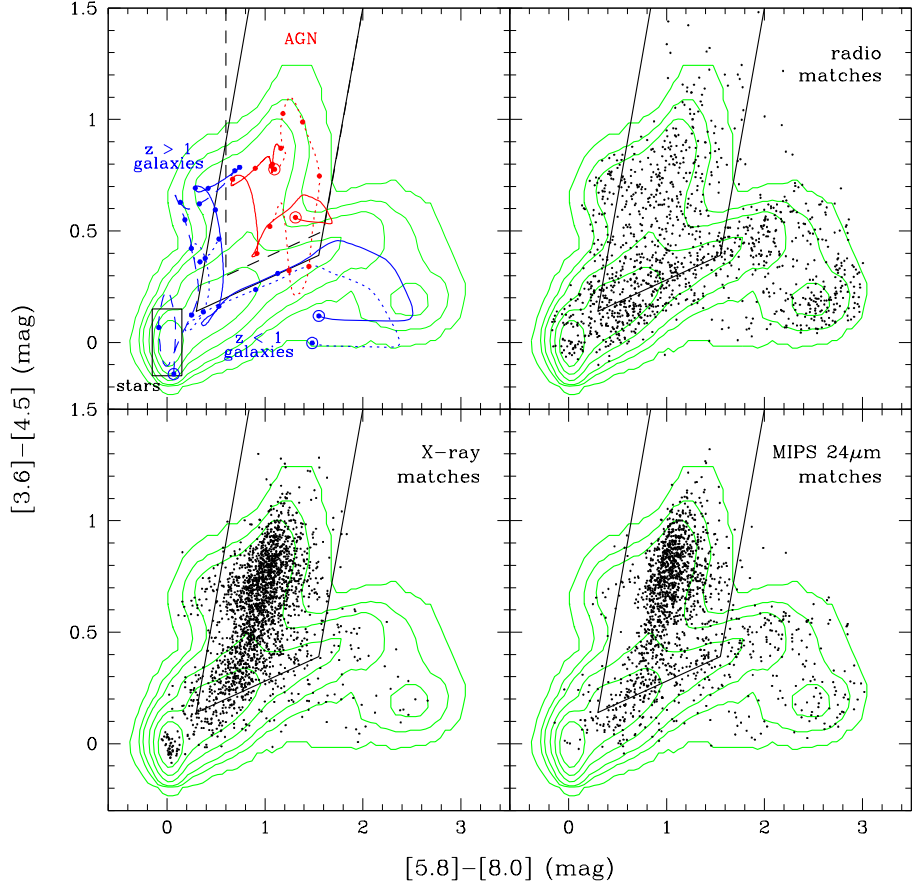}
\caption{SDWFS mid-IR color-color diagrams for the Bo{\"o}tes field. The green contours show the distribution
of all 39,522 Photometry Group 3 sources. The objects are binned in 0.05 mag bins in both axes, and the contours are drawn at levels
of 2, 10, 20, 50, 100, and 150 objects per bin, counting from the outer contour. Top-left panel:
black dashed and solid line ``wedges'' show the
original \cite{2005ApJ...631..163S} AGN selection region and the modified AGN region we define to encompass the distribution
of XBo\"otes X-ray sources discussed in \cite{2008ApJ...679.1040G}.  The black box at color zero is the 
region occupied by normal stars and also nearby galaxies without PAH emission.
The blue solid, dotted, and dashed lines show the evolution of the mid-IR colors of late-type spiral, irregular, and early-type galaxies with 
redshift based on the SED models of \cite{2010ApJ...713..970A}.  The red dotted and solid lines show the evolution
of a pure AGN and a composite with equal contributions from the AGN and its late-type host to the bolometric luminosity.
The dots on each track mark redshift increments of 0.5 (1.0) from $z=0$ to 3 (to 6) for galaxies (AGNs) that start at the $z=0.0$ bulls eye.
Other panels: Photometry Group 3 mid-IR counterparts of radio sources (top right), X-ray sources (bottom left), 
and MIPS 24 \micron~AGN candidates (bottom right) are shown.
\label{fig:colcolintro}}
\end{figure}

\begin{figure}
\centering
\includegraphics[width=16cm]{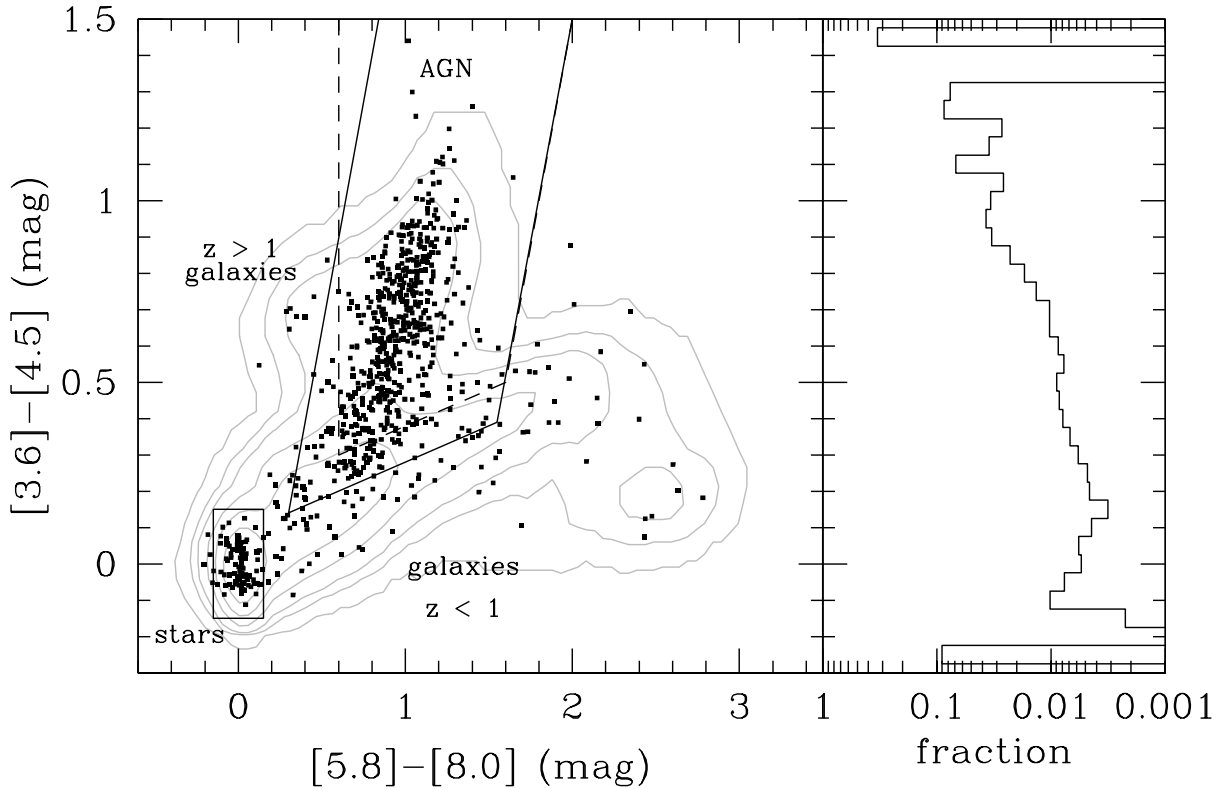}
\caption{
Left panel: mid-IR color-color distribution of all Photometry Group 3, Variability Level 2 sources in SDWFS. The contours show
the distribution of all Photometry Group 3 sources, as described in Figure~\ref{fig:colcolintro}. The \cite{2005ApJ...631..163S} 
AGN selection region is shown by the dashed wedge, and the modified AGN wedge is shown with the solid line. 
The stellar box encompasses sources with (stellar) colors near zero. The majority of mid-IR variable objects 
(76\%) are AGNs.
Also 11\% of objects in the stellar box are variable. Right panel: fraction of Photometry Group 2 sources (e.g., sources lacking reliable
$[5.8]$ and/or $[8.0]$ photometry) that are Variability Level 2 variables as a function of $[3.6]-[4.5]$ color. 
\label{fig:colcol_hist}}
\end{figure}

\begin{figure}
\centering
\includegraphics[width=16cm]{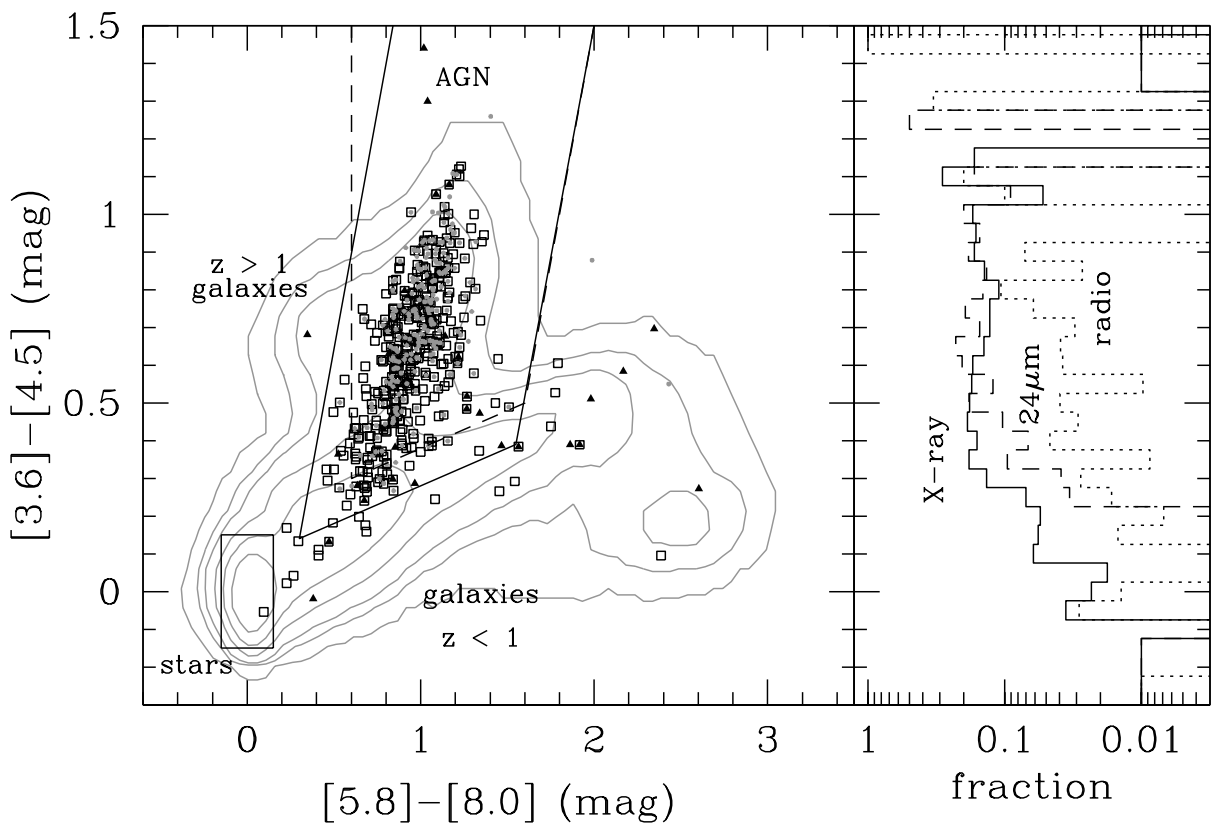}
\caption{
Left panel: mid-IR color-color distribution of the Photometry Group 3, Variability Level 2 sources selected as AGNs based on
 their X-ray (squares), radio (triangles), or 24 \micron~emission (dots).  
The contours and selection regions are the same as in Figure~\ref{fig:colcol_hist}.
Right panel: histograms of the fraction of the Photometry Group 2 X-ray (solid), 24 \micron~(dashed), and radio (dotted) sources identified
as Variability Level 2 variables as a function of  $[3.6]-[4.5]$ color.
\label{fig:col-col_QSO}}
\end{figure}

\begin{figure}
\centering
\includegraphics[width=16cm]{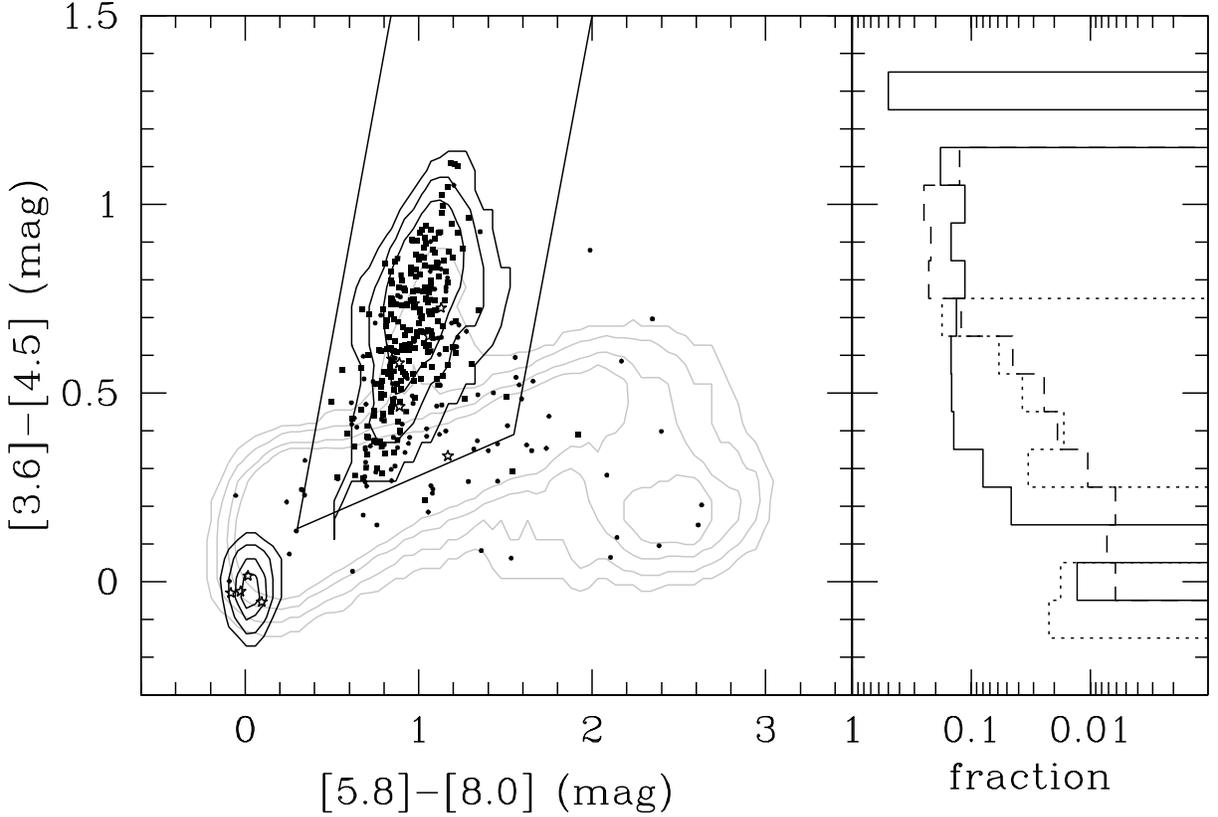}
\caption{
Left panel: mid-IR color-color distribution of the Photometry Group 3, Variability Level 2 sources with spectroscopic redshifts
from AGES.  Stars, circles, and squares correspond to objects with stellar, galaxy, or AGN spectroscopic template
matches.  The contours show the distribution of galaxies (gray), stars, and AGNs with AGES redshifts.
Right panel: fraction of objects with stellar, galaxy, or AGN template matches (dotted, dashed, and solid line, respectively) 
that are the Photometry Group 2, Variability Level 2 variables as a function of  $[3.6]-[4.5]$ color. 
\label{fig:col-col_SDSS}}
\end{figure}

\begin{figure}
 \centering
 \includegraphics[width=16cm]{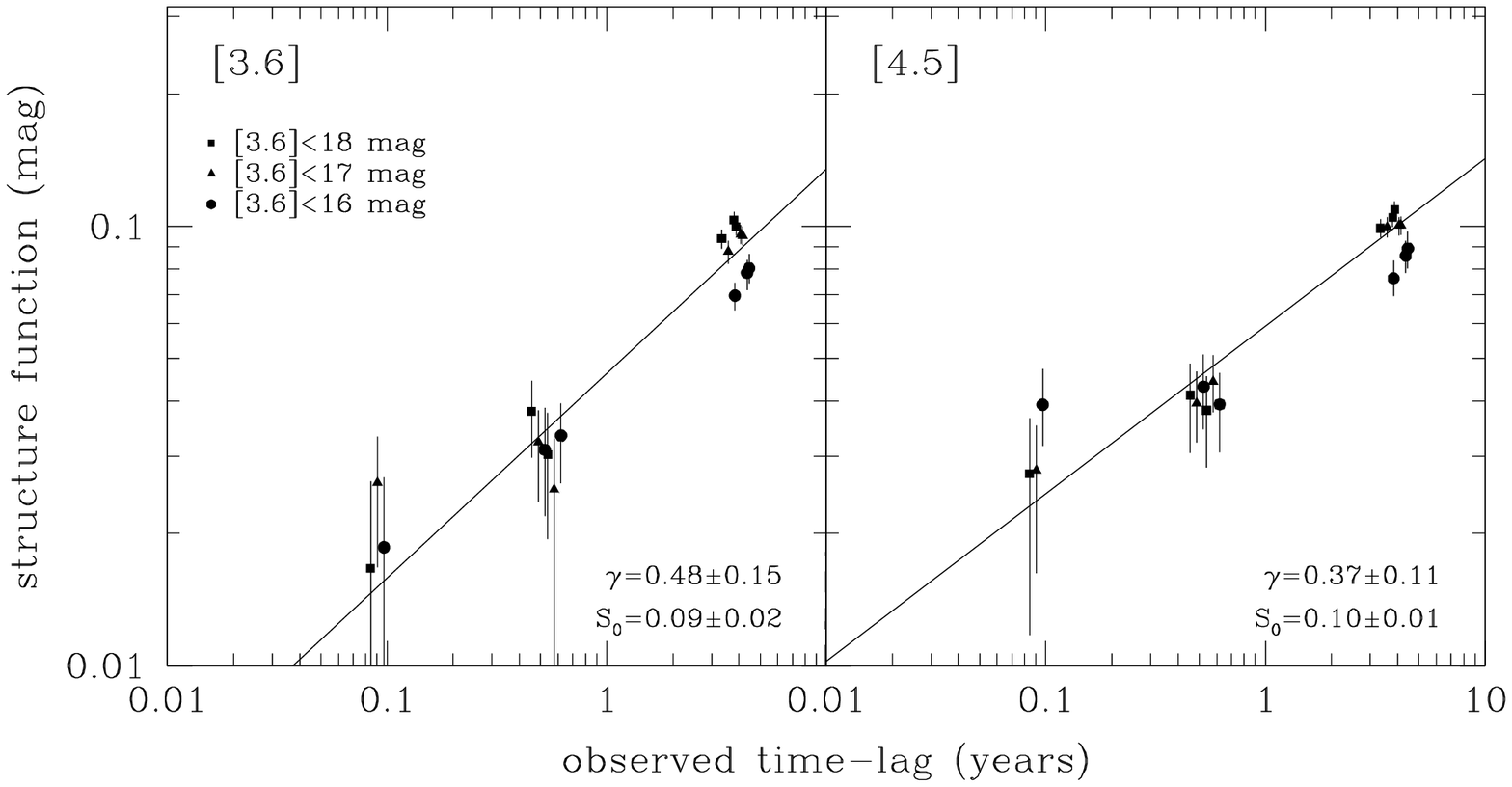}\\
 \vspace{4mm}
 \includegraphics[width=16cm]{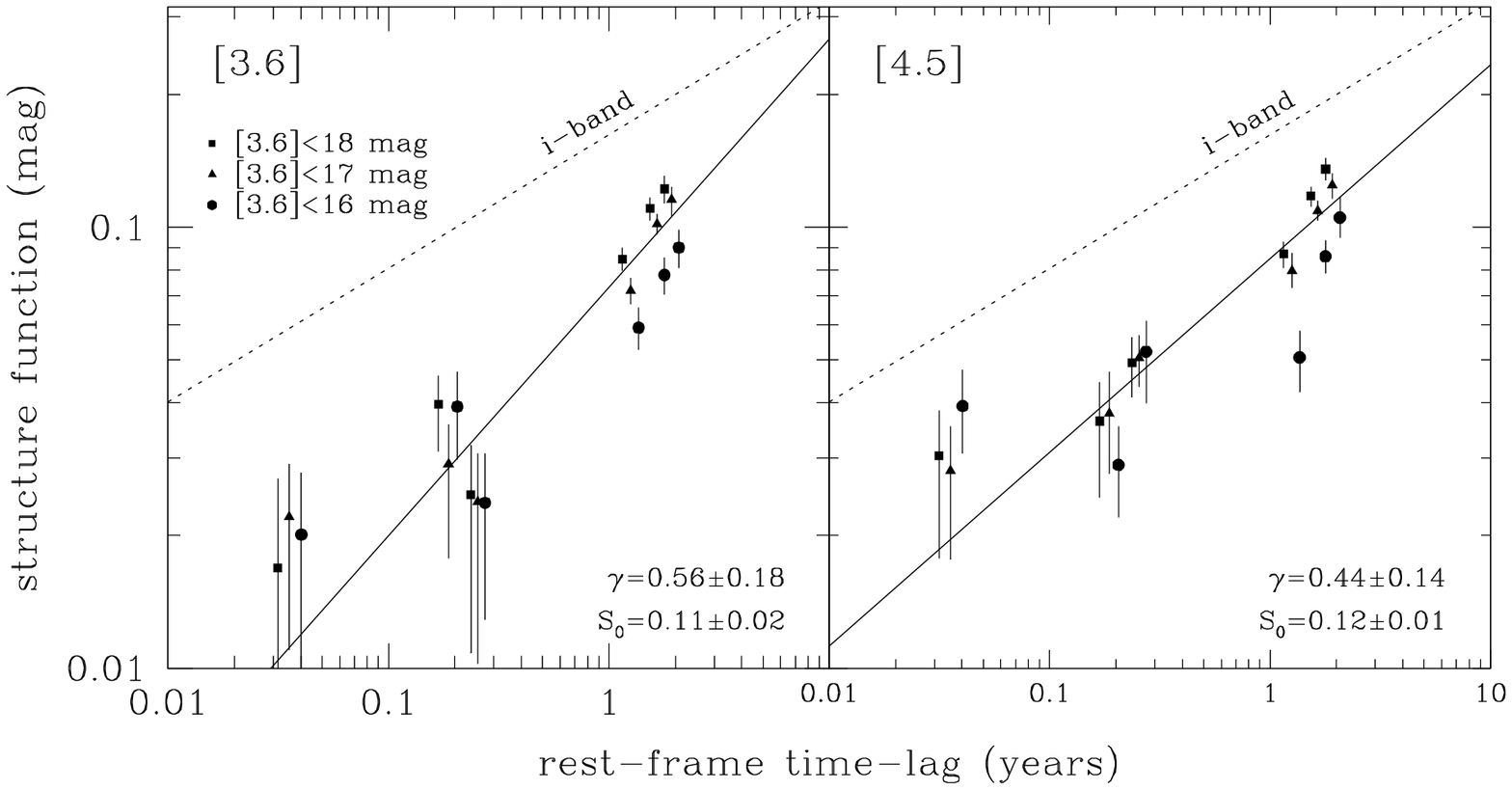}
 \caption{Mid-IR AGN structure functions (Equation~(\ref{eqn:sf})) in the observed frame (top) and rest-frame (bottom). 
They are based on $\sim180$, 690, and 1100 $z>1$ SDWFS objects with $[3.6]<16$ mag (dots), 
$[3.6]<17$ mag (triangles), and $[3.6]<18$ mag (squares), respectively. We also show the {\it i}-band
rest-frame structure function from \cite{2004ApJ...601..692V} in the lower panels.
 \label{fig:sf_obs}}
 \end{figure}

\begin{figure}
 \centering
 \includegraphics[width=16cm]{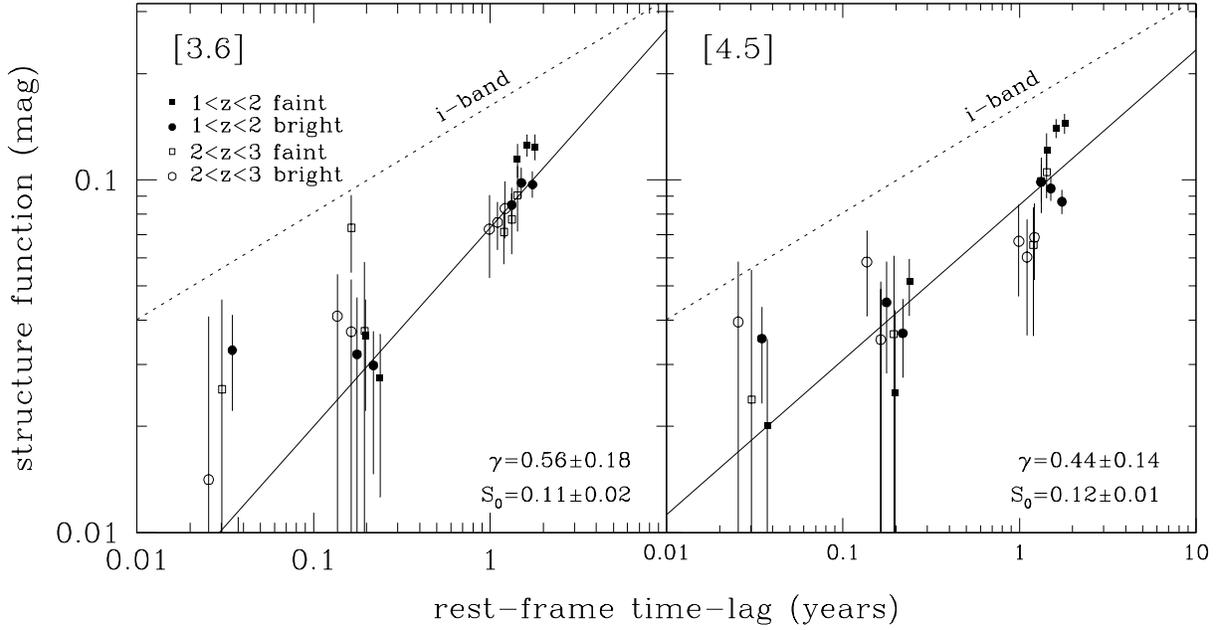}\\
 \caption{Rest-frame mid-IR AGN structure functions for the objects with $[3.6]<18$ mag.
 We divide the objects into two redshift bins, $1<z<2$ (filled symbols) and $2<z<3$ (open symbols), and also into
two brightness levels, objects brighter (circles) and fainter (squares) than the median for the redshift bin. 
For time lags of half a year and longer, fainter and lower redshift sources show greater variability.  
The solid lines are the mid-IR structure function fits from the bottom panel of Figure~\ref{fig:sf_obs}, and
the dotted lines are the {\it i}-band rest-frame structure functions from \cite{2004ApJ...601..692V}.
 \label{fig:sf_bin}}
 \end{figure}

Given the structure function, we can now estimate the
expected statistical properties of the AGNs in the current survey or any extension 
by adding new epochs.  The structure function $S(\tau)$
is related to the correlation function by $C(\tau) = V_\infty^2-S(\tau)$,
where $V_\infty$ is the variance as $\tau\rightarrow \infty$. This
makes it straightforward to compute the mean $\chi^2$ of an AGN
$    \langle \chi^2 \rangle =
  \langle (m_i-\langle m \rangle)^2/\sigma^2
$
since $\langle m_i m_j \rangle = \sigma^2 \delta_{ij}+ C(|t_i-t_j|)$, where $\sigma$ is the measurement uncertainty.
Because we fit for the mean, $\langle m\rangle$, the exact value of $V_\infty$ is irrelevant. 
For our observed frame power-law structure
function, the result for the SDWFS data and cadence is
$\langle \chi^2 \rangle = 3 + 1.6 S_0^2/\sigma^2$, where
$S = S_0 (\tau/\tau_0)^{0.52}$, $S_0\simeq 0.1$ mag, and $\sigma$ is the typical photometric uncertainty. 
The constant $3$ is simply the number of degrees of freedom after
fitting the mean, so a non-variable source would have $\langle \chi^2 \rangle = 3$.  
This expression makes it clear why our
AGN completeness is so low and drops rapidly with magnitude (increasing $\sigma$, see Figure~\ref{fig:var3645}).

We can, in fact, simulate the completeness. We can decompose the
correlation matrix $C_{ij} = C(|t_i-t_j|)$ into
eigenvectors ${\bf v}_i$ and eigenvalues $\lambda^2_i$.
If $G(x)$ is a Gaussian deviate of dispersion $x$, then
random realizations of light curves consistent with the
structure function are
\begin{equation}
{\bf m } = \sum_i {\bf v}_i G(\lambda_i) + G(\sigma)
\end{equation}
(see \citealt{1992ApJ...398..169R}). If we
assume the variability in both bands is identical but the
noise is independent, then we can compute the fraction of
such AGNs we would select as a function of the noise $\sigma$
using Monte Carlo simulations.
Figure~\ref{fig:compl} shows the approximate completeness as a function of
magnitude for sources with $r>0.8$ and either $\sigma_{12}>2$
or $4$.  We use the observed structure function and the normalized
measurement uncertainties to track the median values for [3.6] in
Figure~\ref{fig:var3645}.  For simplicity, the [4.5] band is treated as being
identical to the [3.6] band.  As might be expected from the
structure of Figure~\ref{fig:var3645}, the completeness is high when the 
measurement errors are small compared to the amplitude of the structure
function and then drops rapidly for fainter sources.

In general, the statistics of these simulated light curves agree with observations,
indicating that our estimates of the structure function are correct. However, they
also show that we are ``pushing the envelope'' of variability, because for most AGNs the 
variability amplitudes are merging into the measurement uncertainties, as can be seen in Figure~\ref{fig:var3645}.


\section{Summary}
\label{sec:summary}

In this paper, we performed DIA photometry on the four epochs of SDWFS data.
The common area of the {\it Spitzer} mosaics covers 8.1 deg$^2$ of 
the NDWFS Bo{\"o}tes field, and contains $474,179$ mostly extragalactic objects. 
We concentrate on the deeper 3.6 and 4.5 \micron~channels, and provide
variability catalogs and light curves. These variability data are cross-correlated 
with the AGES redshift survey, X-ray and radio data, and 24 \micron~AGN candidates 
from the same area of the sky to study the variability of various classes of objects. 

We study the mid-IR variability of objects based on the significance level of their 
joint variability ($\sigma_{12}$) in both the [3.6] and [4.5] channels focusing on the sources
that also show a high correlation ($r>0.8$) between the two channels.
Our standard selection criteria with $\sigma_{12}>2$ identify $\sim5100$ objects as variables, which 
constitute 1.1\% of all detected sources in the field. We find that the majority (76\%) 
of the mid-IR variable sources in the NDWFS Bo{\"o}tes field are AGNs,
where we define AGNs to be objects that satisfy any of the following criteria: (1) lie inside the modified AGN wedge,
(2) are X-ray, 24 \micron~or radio AGN candidates, (3) have $z>1$, or (4) 
have an AGN spectroscopic template in any region. 

Amongst all the extragalactic classes of objects, AGNs are the primary sources able to 
produce significant changes in their mid-IR luminosity over a several year timescale. Thus, mid-IR variability may be used as a  
new tool for selecting AGNs, especially if combined with other methods. However, in our relatively 
shallow few-epoch survey, only $\sim$ 15\% of AGNs are sufficiently variable to be detected; therefore, 
variability-selected samples of AGNs are highly incomplete.
This incompleteness is caused by the typical variability amplitude being comparable to the survey 
photometric errors, and is not due to the small number of available epochs.
For our standard criteria, 14\% of X-ray sources, 17\% of 24 \micron\ selected AGN candidates, and 
15\% of spectroscopically confirmed AGNs are detected by their variability. 

The variable AGNs primarily occupy the mid-IR AGN selection wedge of \cite{2005ApJ...631..163S}, 
with an extension to bluer $[3.6]-[4.5]$ colors where the host galaxy dominates the mid-IR colors 
of low-luminosity AGNs (see \citealt{2008ApJ...679.1040G,2010ApJ...713..970A}).
If we redefine the original \cite{2005ApJ...631..163S} selection wedge 
to encompass X-ray or variability-selected AGNs,
we find that variable objects bluer than $[3.6]-[4.5]>0.3$ mag are likely to be AGNs. This opens a new window
for the {\it Spitzer} Warm Mission, where the $[3.6]-[4.5]>0.3$ mag color cut combined 
with the variability criterion will be a solid indication of AGN activity.

The structure function of $z>1$ AGNs is well described by a power law with a logarithmic slope of
$\gamma = 0.56 \pm 0.18$ and amplitude $S_0=0.11\pm0.02$ mag in [3.6] and
$\gamma = 0.44 \pm 0.14$, $S_0=0.12\pm0.01$ mag in [4.5] for
a rest-frame time lag of $\tau_0=2$ years.
These structure functions are steeper than those observed in optical bands, and we argue that 
this is due to the inability of either disk or dust emission to respond on the 
short timescales seen in optical studies. 
The mid-IR structure functions go approximately as $\tau^{1/2}$, which corresponds to  
the short timescale ($\tau \ll \tau_{\rm model}$) behavior in the damped random walk
model used by \cite{2009ApJ...698..895K}, \cite{2010ApJ...708..927K}, and \cite{2010arXiv1004.0276M}. 
If we examine the variability of various subsamples, we detect more variability in low 
luminosity objects observed at shorter rest-frame wavelengths. 
The mid-IR sources corresponding to X-ray and 24 \micron~AGN have structure functions identical to
that of $z>1$ AGNs. However, this is not the case for the counterparts of radio sources which have $S_0\simeq0.05\pm0.01$ 
mag in both [3.6] and [4.5] bands, as compared to $S_0\simeq0.10\pm0.01$ for either $z>1$ objects, X-ray, or MIPS sources. 
One explanation may be that the radio sample has more contamination by non-AGN (e.g., due to confusion between radio cores and lobes), 
another is that the radio-selected AGNs are accreting in a different mode (lower 
accretion rates or efficiencies; e.g., \citealt{2008A&A...490..893T}) and 
that variability in such a mode is inherently lower.

\begin{figure}
 \centering
 \includegraphics[width=16cm]{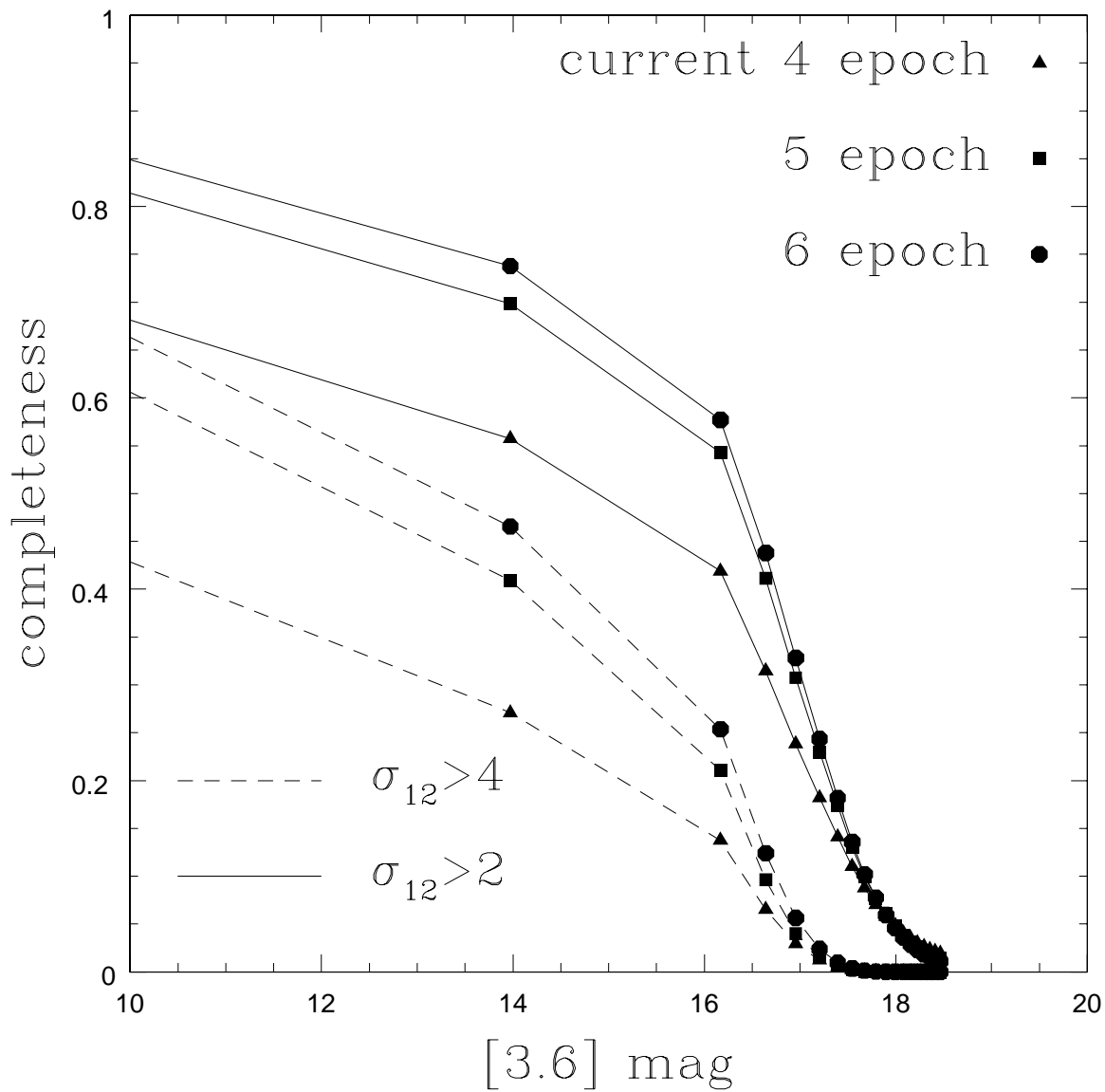}\\
 \caption{Detection completeness for variable AGNs as a function of magnitude for sources
 with high correlation $r>0.8$ and either $\sigma_{12}>2$ or $4$ (Variability Level 2 or 4). We show the completeness
for the current four epoch SDWFS survey, for the current survey plus one epoch in 2011 March (five epochs), and
for the current survey plus two epochs in 2011 and 2012 March (six epochs).
 \label{fig:compl}}
 \end{figure}

Expanding the SDWFS survey with new epochs during the {\it Spitzer} Warm Mission would represent a significant improvement.
Given our structure functions, we can estimate the results from
expanding the SDWFS survey with new epochs. We considered two cases of (1) doing one new
epoch in 2011 March and (2) doing new epochs in both 2011 and
2012 March. The improvements in the mean $\chi^2$ are significant, with
$\langle \chi^2 \rangle = 4 + 3.0 S_0^2/\sigma^2$ and $5 + 4.3 S_0^2/\sigma^2$, respectively, for AGNs,
compared to $\langle \chi^2 \rangle = 4$ or 5 for non-variable objects.
These represent a significant improvement over the SDWFS baseline, where $\langle \chi^2 \rangle = 3 + 1.6 S_0^2/\sigma^2$ for AGN 
and 3 for non-variable objects.
Figure~\ref{fig:compl} shows the increases in
completeness for these five and six epoch scenarios.  Most of the
gain for detecting variable sources comes from adding one additional epoch.
The greatest astrophysical gain from two additional epochs comes from being
able to better map out and subdivide the structure function by
luminosity, redshift, and source type to better characterize the
physics of AGN variability at these wavelengths.
In particular, it would confirm the different slope of the mid-IR structure 
function from that observed in the optical.


\acknowledgments
This work is based on observations made with the \sst, 
which is operated by the Jet Propulsion Laboratory,
California Institute of the Technology under contract
with the National Aeronautics and Space Administration (NASA).
Support for this work was provided by NASA through award numbers 1310744 
(C.S.K. and S.K.), 1314516 (M.L.N.A.) and 1317692 (H.A.S.) issued by JPL/CalTech.
CSK \& SK are also supported by NSF grant AST-0708082.
The NDWFS and the research of A.D. and B.T.J. are supported by 
the National Optical Astronomy Observatory, which is operated by the Association 
of Universities for Research in Astronomy (AURA) under cooperative agreement 
with the National Science Foundation. Support for M.B. was provided by 
the W. M. Keck Foundation.


\newpage


\clearpage

\begin{deluxetable}{ccccccc|cccccc|c}
\tabletypesize{\scriptsize}
\rotate
\tabletypesize{\tiny}
\tablecaption{SDWFS-DIA Variability Catalog.\label{tab:SDWFSvar}}
\tablewidth{0pt}
\tablehead{
\colhead{Object Name} & \colhead{R.A.} & \colhead{Decl.} & \colhead{[3.6]} &
\colhead{[4.5]} & \colhead{[5.8]} & \colhead{[8.0]} &
\colhead{$N\epsilon$} & \colhead{$v([3.6])$} & \colhead{$v([4.5])$} &
\colhead{$C_{12}$} & \colhead{$r$} & \colhead{$\sigma_{12}$} & \colhead{Mask} \\
\colhead{} & \colhead{(deg)} & \colhead{(deg)} & \colhead{(mag)} & 
\colhead{(mag)} & \colhead{(mag)} & \colhead{(mag)} &
\colhead{} & \colhead{(mag)} & \colhead{(mag)} &
\colhead{(mag$^2$)} & \colhead{} & \colhead{} & \colhead{} 
}
\startdata
SDWFS J143040.36+322721.8 & 217.668148 & 32.456048 & 17.666 & 17.488 & 16.792 & 16.070 & 4 &  0.051 &  0.062 &  0.001 &  0.178 &  0.894 & 1 \\
SDWFS J143034.08+322729.0 & 217.642013 & 32.458058 & 17.937 & 17.466 & 17.142 & 16.928 & 4 &  0.067 &  0.185 &  0.002 &  0.127 &  0.758 & 1 \\
SDWFS J143033.71+322730.3 & 217.640453 & 32.458420 & 18.725 & 18.352 & 17.576 & 16.869 & 4 &  0.181 &  0.223 &  0.029 &  0.718 &  0.392 & 1 \\
SDWFS J143047.11+322729.5 & 217.696285 & 32.458194 & 17.121 & 16.995 & 16.898 & 14.735 & 4 &  0.036 &  0.096 &  0.002 &  0.511 &  0.295 & 1 \\
SDWFS J142951.47+322726.0 & 217.464455 & 32.457230 & 18.361 & 17.875 & 17.538 & 16.447 & 4 &  0.135 &  0.102 &  0.009 &  0.627 &  0.734 & 1 \\
SDWFS J143015.91+322733.8 & 217.566284 & 32.459402 & 19.732 & 20.200 & 18.262 & 99.999 & 1 & 99.999 & 99.999 & 99.999 & 99.999 & 99.999 & 1 \\
\enddata
\tablecomments{This is an abridged version of the full table, presenting the variability statistics for the combined channels [3.6] and [4.5] only.
The electronic table includes additional columns showing the variability statistics 
for the separate channels [3.6] and [4.5], and the photometric uncertainties for the IRAC channels. Here, quantities $v([X])$ and $N\epsilon$, are respectively, 
the variability standard deviation in channel $[X]$ and the common number of epochs in both channels used in the calculation of the joint variability. 
With $C_{12}$, $r$, and $\sigma_{12}$ we denote the covariance and the correlation between the channels 1 and 2, and the joint variability significance, 
respectively. The mask is 0 (1) if there is (is no) bright star with the
{\sc2mass} $K$-band magnitude within $260(1-K/11)$ arcsec affecting the measurement.
The Vega magnitudes presented in the table are measured in 4 arcsec apertures 
(see Ashby et al. 2009).  The error code for magnitudes, reflecting no 
measurement, is 99.999. The error code for $v$, $C_{12}$, $r$, and $\sigma_{12}$ is 99.999. \\ 
(This table is  available in its entirety in a machine-readable form in the online journal. 
A portion is shown here for guidance regarding its form and content.)} 
\end{deluxetable}

\begin{deluxetable}{lc|cccc|cccc}
\tabletypesize{\scriptsize}
\tablecaption{The Light Curves for the SDWFS-DIA Variability Catalog.\label{tab:SDWFSlc}}
\tablewidth{0pt}
\tablehead{
\multicolumn{2}{c}{} &
\multicolumn{4}{c}{Channel 1} &
\multicolumn{4}{c}{Channel 2} \\
R.A. &
Decl. & 
Epoch 1 &
Epoch 2 &
Epoch 3 &
Epoch 4 &
Epoch 1 &
Epoch 2 &
Epoch 3 &
Epoch 4 \\
(deg) &
(deg) &
(mag) &
(mag) &
(mag) &
(mag) &
(mag) &
(mag) &
(mag) &
(mag) \\
}
\startdata
217.668148 & 32.456048 & 17.703 & 17.823 & 17.741 & 17.771 & 17.457 & 17.518 & 17.547 & 17.409 \\
217.642013 & 32.458058 & 17.894 & 18.027 & 17.881 & 17.952 & 17.290 & 17.455 & 17.566 & 17.728 \\
217.640453 & 32.458420 & 19.037 & 18.793 & 18.769 & 18.598 & 18.721 & 18.220 & 18.288 & 18.371 \\
217.696285 & 32.458194 & 17.107 & 17.128 & 17.132 & 17.191 & 17.019 & 16.995 & 17.195 & 17.143 \\
217.464455 & 32.457230 & 18.385 & 18.070 & 18.166 & 18.148 & 17.925 & 17.780 & 17.696 & 17.876 \\
217.566284 & 32.459402 & 20.055 & 19.751 & 20.271 & 19.649 & 99.999 & 99.999 & 19.692 & 99.999 \\
\enddata
\tablecomments{The error code for magnitudes, reflecting no measurement, is 99.999.\\
(This table is available in its entirety in a machine-readable form in
the online journal. A portion is shown here for guidance regarding its
form and content.)}
\end{deluxetable}

\begin{deluxetable}{ccccccc}
\rotate
\tabletypesize{\scriptsize}
\tablecaption{Mid-IR Variable Objects in the SDWFS.\label{tab:resultsSDWFS}}
\tablewidth{0pt}
\tablehead{
\makebox[2.0cm][c]{Variability} & 
\makebox[2.5cm][c]{Variability} & 
\makebox[2.5cm][c]{Old AGN Wedge} & 
\makebox[2.5cm][c]{Modified AGN Wedge} & 
\makebox[2.0cm][c]{Stellar Box} & 
\makebox[2.0cm][c]{Outside$^a$} & 
\makebox[2.0cm][c]{All Abjects$^a$}\\
Level & $\sigma_{12}$ & & & & &  
}
\startdata
\multicolumn{7}{c}{Number (fraction/false positive rate$^{b}$) of variable objects: Photometry Group 1} \\
\cline{1-7}
 Total & ...  & 43468            & 76005              & 45939            & 144263               & 474179 \\
2      & $>2$ & 746 (1.7\%/28\%) &  1015 (1.3\%/33\%) & 555 (1.2\%/58\%) & 832 (0.6\%/71\%)    & 5107 (1.1\%/57\%) \\
3      & $>3$ & 397 (0.9\%/17\%) &   501 (0.7\%/22\%) & 226 (0.5\%/54\%) & 326 (0.2\%/56\%)    & 2208 (0.5\%/47\%) \\
4      & $>4$ & 246 (0.6\%/11\%) &   293 (0.4\%/17\%) & 102 (0.2\%/51\%) & 156 (0.1\%/53\%)    & 1071 (0.2\%/44\%) \\
\\
\multicolumn{7}{c}{Number (fraction/false positive rate$^{b}$) of variable objects: Photometry Group 2} \\
\cline{1-7}
 Total & ...  &  29181            & 54920            & 18150            & 113299              & 213594 \\
2      & $>2$ &  598 (2.0\%/18\%) & 774 (1.4\%/24\%) & 133 (0.7\%/65\%) & 512 (0.5\%/74\%)    & 1557 (0.7\%/50\%) \\
3      & $>3$ &  340 (1.2\%/9\%)  & 405 (0.7\%/13\%) &  58 (0.3\%/56\%) & 166 (0.1\%/56\%)    &  668 (0.3\%/30\%) \\
4      & $>4$ &  219 (0.8\%/5\%)  & 250 (0.5\%/9\%)  &  24 (0.1\%/42\%) &  71 (0.1\%/55\%)    &  359 (0.2\%/24\%) \\
\\
\multicolumn{7}{c}{Number (fraction/false positive rate$^{b}$) of variable objects: Photometry Group 3} \\
\cline{1-7}
 Total & ...  &  6677            & 12741           & 7567             & 19214            & 39522 \\
2      & $>2$ &  496 (7.4\%/6\%) & 584 (4.6\%/7\%) &  86 (1.1\%/58\%) & 105 (0.5\%/44\%) & 775 (2.0\%/19\%) \\
3      & $>3$ &  310 (4.6\%/3\%) & 349 (2.7\%/4\%) &  44 (0.6\%/48\%) &  35 (0.2\%/44\%) & 428 (1.1\%/12\%) \\
4      & $>4$ &  204 (3.1\%/2\%) & 224 (1.8\%/3\%) &  21 (0.3\%/33\%) &  15 (0.1\%/30\%) & 260 (0.7\%/9\%)
\enddata
\tablecomments{Photometry Groups 1 and 2 are defined in Section~\ref{sec:definitions}; only 49\% and 86\%
of objects in these groups have all four IRAC band measurements, respectively. The remaining objects lack
either [5.8],  [8.0], or both measurements, so we are unable to place them on the mid-IR 
color-color diagram (i.e., we were unable to tell if they fall into the modified/old AGN wedge or stellar box). 
By definition, Photometry Groups 3 includes only objects with all four IRAC measurements.\\
$^a$ Column ``Outside'' includes objects with all four IRAC band measurements only
that are both outside of the ``Modified AGN Wedge'' and the ``Stellar Box.'' Column ``All Objects'' contains all objects,
regardless of the number of IRAC band measurements.\\
$^{b}$ The fraction of variable sources is a ratio of the number of variable objects above a given 
level of $\sigma_{12}$ and $r>0.8$ to the total number of objects in a respective region. The false positive rate is 
estimated from the number of sources above the same level of $\sigma_{12}$ but with $r<0.5$ (see Section~\ref{sec:definitions} and Figure~\ref{fig:covariance}).}
\end{deluxetable}

\clearpage
 
\begin{deluxetable}{cccccc}
\tabletypesize{\scriptsize}
\tablecaption{Mid-IR Variability of Photometrically-Selected AGNs.\label{tab:resultsAGES}}
\tablewidth{0pt}
\tablehead{
\makebox[2.0cm][c]{Variability} & 
\makebox[2.5cm][c]{Variability} &
\makebox[2.5cm][c]{MIPS QSOs} & 
\makebox[2.5cm][c]{X-ray} & 
\makebox[2.5cm][c]{FIRST} &
\makebox[2.5cm][c]{WSRT}\\
Level & $\sigma_{12}$ & & & &
}
\startdata
& & \multicolumn{4}{c}{Number (fraction/false positive rate) of variable objects: Photometry Group 1} \\
\cline{1-6}
Total & ...  & 1964             & 3231             &  313             & 2046 \\
2     & $>2$ & 230 (11.7\%/5\%) & 392 (12.1\%/4\%) &   12 (3.8\%/8\%) & 48 (2.3\%/13\%) \\
3     & $>3$ & 151 (7.7\%/3\%)  & 253 (7.8\%/2\%)  &    8 (2.6\%/3\%) & 27 (1.3\%/8\%) \\
4     & $>4$ & 107 (5.4\%/2\%)  & 164 (5.1\%/1\%)  &    4 (1.3\%/1\%) & 17 (0.8\%/4\%) \\
\\
& & \multicolumn{4}{c}{Number (fraction/false positive rate) of variable objects: Photometry Group 2} \\
\cline{1-6}
Total & ...  & 1942             & 2950             &  292             & 1883 \\
2     & $>2$ & 229 (11.8\%/5\%) & 391 (13.3\%/3\%) &   12 (4.1\%/7\%) & 48 (2.5\%/10\%) \\
3     & $>3$ & 151 (7.8\%/3\%)  & 253 (8.6\%/2\%)  &    8 (2.7\%/3\%) & 27 (1.4\%/6\%) \\
4     & $>4$ & 107 (5.5\%/2\%)  & 164 (5.6\%/1\%)  &    4 (1.4\%/1\%) & 17 (0.9\%/3\%) \\
\\
& & \multicolumn{4}{c}{Number (fraction/false positive rate) of variable objects: Photometry Group 3} \\
\cline{1-6}
Total & ...  & 1637             & 2217             &  195             & 1233 \\
2     & $>2$ & 225 (13.7\%/4\%) & 374 (16.9\%/2\%) &   11 (5.6\%/4\%) & 43 (3.5\%/9\%) \\
3     & $>3$ & 151 (9.2\%/3\%)  & 245 (11.1\%/1\%) &    8 (4.1\%/1\%) & 25 (2.0\%/5\%) \\
4     & $>4$ & 107 (6.5\%/1\%)  & 160 (7.2\%/1\%)  &    4 (2.1\%/1\%) & 15 (1.2\%/4\%) \\
\enddata
\tablecomments{The fraction of variable sources is a ratio of the number of variable objects above a given 
level of $\sigma_{12}$ and $r>0.8$ to the total number of objects in a respective class. The false positive rate is 
estimated from the number of sources above the same level of $\sigma_{12}$ but with $r<0.5$ 
(see Section~\ref{sec:definitions} and Figure~\ref{fig:covariance}).}
\end{deluxetable}


\begin{deluxetable}{ccccc}
\tabletypesize{\scriptsize}
\tablecaption{Mid-IR Variability by AGES Spectroscopic Classification.\label{tab:resultsSpec}}
\tablewidth{0pt}
\tablehead{
\makebox[2.0cm][c]{Variability} & 
\makebox[2.5cm][c]{Variability} &
\makebox[2.5cm][c]{QSOs} & 
\makebox[2.5cm][c]{Stars} & 
\makebox[2.0cm][c]{Galaxies}\\
Level & $\sigma_{12}$ & & & 
}
\startdata
& & \multicolumn{3}{c}{Number (fraction/false positive rate) variable objects: Photometry Group 1} \\
\cline{1-5}
Total & ...  & 2950             & 1121               & 18161 \\
2     & $>2$ & 334 (11.3\%/5\%) &    20 (1.8\%/20\%) & 209 (1.1\%/25\%) \\
3     & $>3$ & 216 (7.3\%/3\%)  &    11 (1.0\%/7\%)  & 108 (0.6\%/15\%) \\
4     & $>4$ & 144 (4.9\%/2\%)  &     6 (0.5\%/5\%)  &  60 (0.3\%/10\%) \\
\\
& & \multicolumn{3}{c}{Number (fraction/false positive rate) of variable objects: Photometry Group 2} \\
\cline{1-5}
Total & ...  & 2820             & 966                & 17405 \\
2     & $>2$ & 333 (11.8\%/4\%) &  18 (1.9\%/14\%)   & 200 (1.1\%/24\%) \\
3     & $>3$ & 216 (7.7\%/2\%)  &  10 (1.0\%/5\%)    & 102 (0.6\%/13\%) \\
4     & $>4$ & 144 (5.1\%/2\%)  &   6 (0.6\%/4\%)    &  58 (0.3\%/8\%) \\
\\
& & \multicolumn{3}{c}{Number (fraction/false positive rate) of variable objects: Photometry Group 3} \\
\cline{1-5}
Total & ...  & 2192             & 491                & 11214 \\
2     & $>2$ & 319 (14.6\%/4\%) &  14 (2.9\%/9\%)    & 175 (1.6\%/17\%) \\
3     & $>3$ & 211 (9.6\%/2\%)  &  10 (2.0\%/3\%)    &  93 (0.8\%/10\%) \\
4     & $>4$ & 142 (6.5\%/1\%)  &   6 (1.2\%/3\%)    &  56 (0.5\%/5\%) \\
\enddata
\tablecomments{The fraction of variable sources is a ratio of the number of variable objects above a given 
level of $\sigma_{12}$ and $r>0.8$ to the total number of objects in a respective class. The false positive rate is 
estimated from the number of sources above the same level of $\sigma_{12}$ but with $r<0.5$ 
(see Section~\ref{sec:definitions} and Figure~\ref{fig:covariance}).}
\end{deluxetable}


\begin{deluxetable}{lccc}
\tabletypesize{\scriptsize}
\tablecaption{Mid-IR AGN Structure Functions\label{tab:sf_fits}}
\tablewidth{0pt}
\tablehead{ & & \multicolumn{2}{c}{Structure Function Parameters}\\
Level & Channel & \makebox[2.5cm][c]{$\gamma$} & \makebox[2.5cm][c]{$S_0$}}
\startdata
\\
\multicolumn{4}{c}{Observed Frame ($\tau_0=4$ years)} \\
\cline{1-4}
$[3.6]<16$ mag & [3.6] & $0.42 \pm 0.14$ &  $0.08 \pm 0.02$ \\
$[3.6]<17$ mag & [3.6] & $0.52 \pm 0.16$ &  $0.09 \pm 0.02$ \\
$[3.6]<18$ mag & [3.6] & $0.52 \pm 0.14$ &  $0.10 \pm 0.02$ \\
\cline{1-4}
$[3.6]<16$ mag & [4.5] & $0.28 \pm 0.11$ &  $0.08 \pm 0.01$ \\
$[3.6]<17$ mag & [4.5] & $0.40 \pm 0.10$ &  $0.10 \pm 0.01$ \\
$[3.6]<18$ mag & [4.5] & $0.45 \pm 0.14$ &  $0.10 \pm 0.01$ \\
\cline{1-4}
\\
\multicolumn{4}{c}{Rest-Frame ($\tau_0=2$ years)} \\
\cline{1-4}
$[3.6]<16$ mag & [3.6] & $0.45 \pm 0.18$ &  $0.08 \pm 0.02$ \\
$[3.6]<17$ mag & [3.6] & $0.66 \pm 0.21$ &  $0.11 \pm 0.02$ \\
$[3.6]<18$ mag & [3.6] & $0.59 \pm 0.14$ &  $0.12 \pm 0.02$ \\
\cline{1-4}
$[3.6]<16$ mag & [4.5] & $0.33 \pm 0.19$ &  $0.09 \pm 0.02$ \\
$[3.6]<17$ mag & [4.5] & $0.44 \pm 0.10$ &  $0.12 \pm 0.01$ \\
$[3.6]<18$ mag & [4.5] & $0.49 \pm 0.11$ &  $0.13 \pm 0.01$ \\
\cline{1-4}

\\
\multicolumn{4}{c}{Rest-Frame ($\tau_0=2$ years)} \\
\cline{1-4}
$1<z<2$ bright & [3.6] & $0.45 \pm 0.17$ &  $0.11 \pm 0.02$ \\
$1<z<2$ faint  & [3.6] & $0.60 \pm 0.14$ &  $0.13 \pm 0.02$ \\
$2<z<3$ bright & [3.6] & $0.30 \pm 0.28$ &  $0.09 \pm 0.02$ \\
$2<z<3$ faint  & [3.6] & $0.19 \pm 0.11$ &  $0.09 \pm 0.01$ \\
\cline{1-4}
$1<z<2$ bright & [4.5] & $0.33 \pm 0.13$ &  $0.10 \pm 0.02$ \\
$1<z<2$ faint  & [4.5] & $0.58 \pm 0.12$ &  $0.15 \pm 0.01$ \\
$2<z<3$ bright & [4.5] & $0.13 \pm 0.15$ &  $0.07 \pm 0.02$ \\
$2<z<3$ faint  & [4.5] & $0.30 \pm 0.20$ &  $0.10 \pm 0.02$ \\
\enddata
\end{deluxetable}


\begin{deluxetable}{lcc}
\tabletypesize{\scriptsize}
\tablecaption{Mid-IR Variability Amplitudes At Fixed $\gamma=0.5$.\label{tab:sf_fits1}}
\tablewidth{0pt}
\tablehead{ & \multicolumn{2}{c}{Amplitude Parameters}\\
Level & \makebox[2.5cm][c]{$S_0$ [3.6] (mag)} & \makebox[2.5cm][c]{$S_0$ [4.5] (mag)}}
\startdata
\\
\multicolumn{3}{c}{$z>1$ objects ($\tau_0=4$ years)} \\
\cline{1-3}
$14<[3.6]<15$ &  $0.07 \pm 0.01$ &  $0.08 \pm 0.01$ \\ 
$15<[3.6]<16$ &  $0.10 \pm 0.01$ &  $0.10 \pm 0.01$ \\
$16<[3.6]<17$ &  $0.10 \pm 0.01$ &  $0.10 \pm 0.01$ \\
$17<[3.6]<18$ &  $0.10 \pm 0.02$ &  $0.09 \pm 0.02$ \\
\cline{1-3}
\\
\multicolumn{3}{c}{X-ray sources inside the modified AGN wedge ($\tau_0=4$ years)} \\
\cline{1-3}
$14<[3.6]<15$ &  $0.11 \pm 0.01$ &  $0.10 \pm 0.01$ \\
$15<[3.6]<16$ &  $0.12 \pm 0.01$ &  $0.10 \pm 0.01$ \\
$16<[3.6]<17$ &  $0.09 \pm 0.01$ &  $0.08 \pm 0.01$ \\
$17<[3.6]<18$ &  $0.10 \pm 0.02$ &  $0.11 \pm 0.03$ \\
\cline{1-3}
\\
\multicolumn{3}{c}{MIPS sources inside the modified AGN wedge ($\tau_0=4$ years)} \\
\cline{1-3}
$14<[3.6]<15$ &  $0.11 \pm 0.01$ &  $0.11 \pm 0.01$ \\
$15<[3.6]<16$ &  $0.10 \pm 0.01$ &  $0.10 \pm 0.01$ \\
$16<[3.6]<17$ &  $0.09 \pm 0.01$ &  $0.09 \pm 0.01$ \\
$17<[3.6]<18$ &  $0.11 \pm 0.03$ &  $0.08 \pm 0.03$ \\ 
\cline{1-3}
\\
\multicolumn{3}{c}{Radio (WSRT$+$FIRST) sources inside the modified AGN wedge ($\tau_0=4$ years)} \\
\cline{1-3}
$14<[3.6]<15$ &  $0.06 \pm 0.02$ &  $0.06 \pm 0.01$ \\
$15<[3.6]<16$ &  $0.04 \pm 0.01$ &  $0.06 \pm 0.02$ \\
$16<[3.6]<17$ &  $0.05 \pm 0.01$ &  $0.05 \pm 0.01$ \\
$17<[3.6]<18$ &  $0.05 \pm 0.01$ &  $0.06 \pm 0.02$ \\ 

\enddata
\end{deluxetable}

\end{document}